\documentclass[fleqn,10pt]{wlscirep}
\usepackage[utf8]{inputenc}
\usepackage[T1]{fontenc}
\usepackage{bm}

\usepackage[utf8]{inputenc}
\usepackage[utf8]{inputenc}
\usepackage{booktabs} 

\newlength{\colonewidth}
\newlength{\coltwowidth}
\newlength{\colthreewidth}
\newlength{\colfourwidth}
\newlength{\colfivewidth}
\newlength{\colsixwidth}
\usepackage{xcolor} 

\title{Building Sustainable and Trustworthy Indigenous Knowledge Preservation Ecosystem}

\author[1,*]{Siguo Bi}
\author[2]{Xin Yuan}
\author[2,*]{Wei Ni}

\affil[1]{Fudan University, Shanghai, China}
\affil[2]{CSIRO, Sydney, Australia}

\affil[*]{e-mail: fdbsg@fudan.edu.cn}

\begin{abstract}
This paper focuses on the essential global issue of protecting and transmitting indigenous knowledge. It reveals the challenges in this area and proposes a sustainable supply chain framework for indigenous knowledge. The paper reviews existing technological solutions and identifies technical challenges and gaps. It then introduces cutting-edge technologies to protect and disseminate indigenous knowledge more effectively. The paper also discusses how the proposed framework can address real-world challenges in protecting and transmitting indigenous knowledge, and explores future research applications of the proposed solutions. Finally, it addresses open issues and provides a detailed analysis, offering promising research directions for the protection and transmission of indigenous knowledge worldwide.
\end{abstract}
\begin{document}

\flushbottom
\maketitle
\thispagestyle{empty}

\noindent \textbf{Key points:} 

\begin{itemize}
\item We analyzed challenges in indigenous knowledge protection and proposed a sustainable supply chain framework integrating knowledge protection and dissemination.

\item We prioritized indigenous data sovereignty in the supply chain by designing a reverse loop to return value to communities and support local livelihoods.

\item We examined key technologies in the supply chain's knowledge collection and dissemination stages, emphasizing the potential of advanced technologies in specific scenarios.

\item We explained how our framework addresses existing challenges in indigenous knowledge protection and transmission, proposing promising technical solutions.

\item We detailed future research directions for indigenous knowledge protection, focusing on security, computational power, resilience, and sustainability.
\end{itemize}

\section{Introduction}

According to the United Nations, there are over 470 million indigenous people worldwide, representing approximately 5,000 distinct cultures. The indigenous peoples are distributed across around 90 countries \cite{un_2024}, including the United States, Canada, Australia, and others. The indigenous peoples play a vital role in global biodiversity and cultural diversity. For instance, the languages of Australia's Aboriginal peoples are integral to their rich cultural heritage. Regrettably, over time, some of these languages have vanished, no longer spoken \cite{Hobson2010,Bromham2021GlobalPO}. The Third National Indigenous Languages Survey (NILS3) highlighted that several languages remain spoken, including some learned by children as their first language \cite{NationalIndigenousLanguagesReport2020}. The survey also identified the languages considered strong, but all are at risk and require urgent measures for preservation \cite{aiatsis_nils3}. Despite this loss, recorded images such as rock art \cite{Stuart2017PigmentCI, nmaFirstRockArt} and videos \cite{firstaustraliansepisode1}, along with audio materials like recordings \cite{nmaSonglines}, have been preserved, offering us invaluable historical testimonies. These records capture the languages, stories, songs, and rituals of the Aboriginal people, serving as crucial windows into understanding and studying their culture. 

In the diverse tapestry of ethnic minorities in southwestern China, the De'ang and Yi peoples, along with the Dai and Hani communities, have developed rich indigenous knowledge for disaster prevention and sustainable agriculture  \cite{BaiYanYing201206009,Dai,Deang,XNZS201510002}. For example, the De'ang people utilize the tradition of phenological calendars, observing natural phenomena to predict climate changes, which is crucial for arranging agricultural activities and also aids in disaster early warning \cite{Deang}.  The Yi people use thick stone slabs to surround the hearth, taking into account the positional relationship between the hearth and the house, as well as the direction of the wind
 \cite{XNZS201510002}. These indigenous knowledge systems reflect the ethnic groups' understanding and strategies for dealing with natural disasters, such as floods, hurricanes, earthquakes, and wildfires \cite{XNZS201510002}. For example, the very recent wildfires in Los Angeles have inflicted devastating harm on society and the economy \cite{cv3unkdf4397r3mvoukg}. These systems also demonstrate the wisdom of these communities in living in harmony with the natural environment \cite{Natalie}.

\begin{figure*}[t]
\centering
\includegraphics[width=13.5cm]  {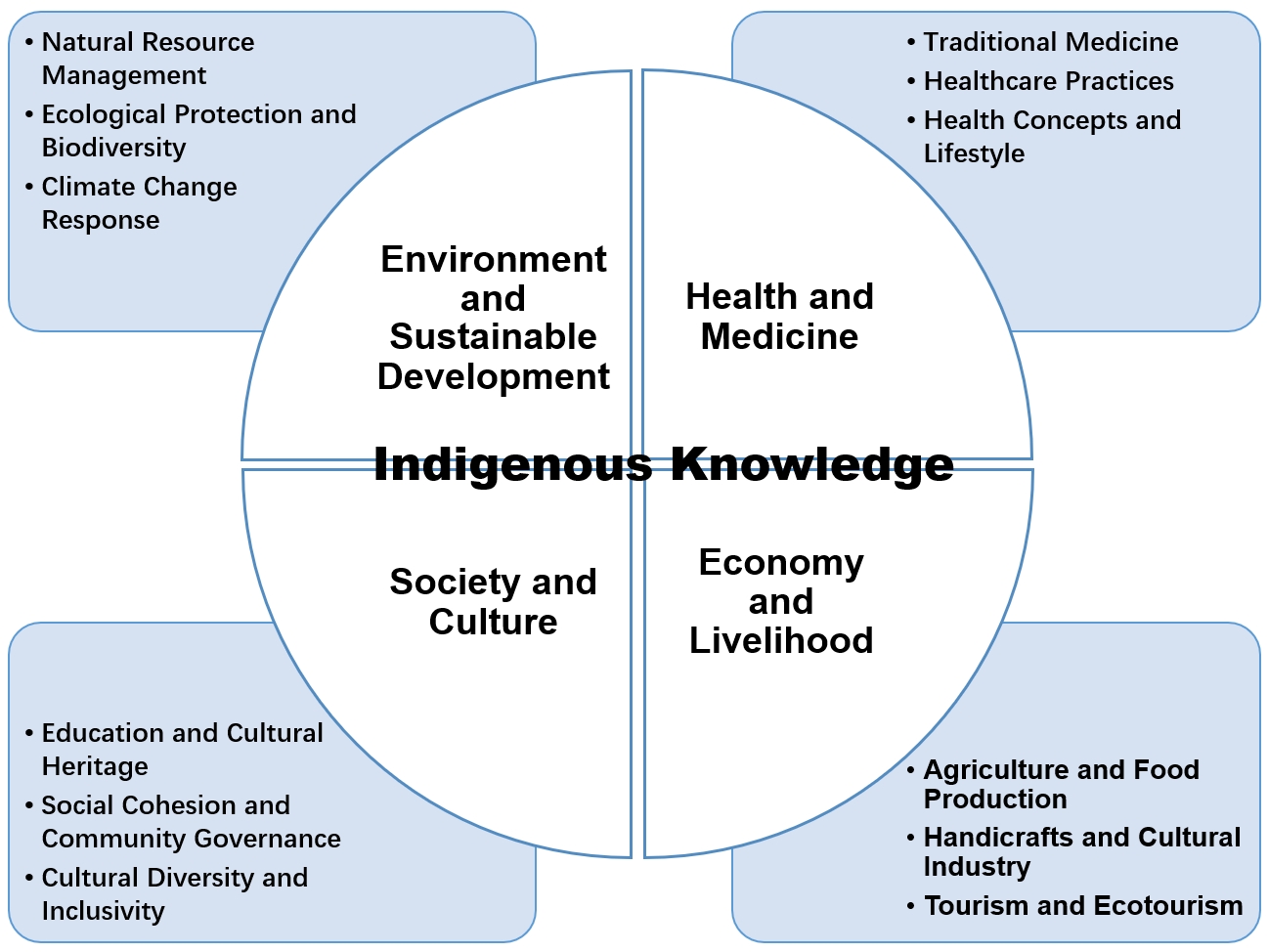}
\caption{\small The potential applications of indigenous knowledge \cite{WIPO1,un_2024_1}.}
\label{F:indigenous}
\end{figure*}

As such, the preservation and dissemination of indigenous knowledge is a critical global endeavor that not only safeguards the rich cultural heritage of indigenous communities but also contributes significantly to the collective global knowledge base \cite{Antonelli2023IndigenousKI}; see Figure \ref{F:indigenous}. The intricate tapestry of human knowledge is enriched by the diverse contributions of indigenous peoples, whose traditions and practices hold invaluable insights into sustainable living, ecological balance, and social cohesion. The Australian government's national cultural policy \cite{Revive2023}, "Revive: A Place with All the Stories, A Story for Every Place," also reflects support for the protection of indigenous cultures and self-determination, aiming to promote the development of indigenous arts and culture through policy means, preserving multiculturalism for future generations. As we recognize the importance of preserving and promoting this wealth of knowledge, it becomes imperative to address the challenges that threaten its existence.

\section{Challenges}
The significance of indigenous knowledge in shaping our understanding of the world cannot be overstated. As we acknowledge the critical role it plays in our global tapestry of wisdom, it is essential to explore the specific challenges that stand in the way of its preservation and dissemination. This exploration will not only shed light on the complexities of these issues but also pave the way for developing strategies to overcome them, ensuring that the invaluable insights of indigenous peoples continue to enrich our collective understanding and contribute to a more sustainable and harmonious world.
The following discussions will delve into four primary concerns that impede the continuity and appreciation of indigenous knowledge: cultural erosion, language loss, intellectual property issues, and the lack of systematic documentation. Each of these areas presents a unique set of obstacles that must be navigated with care and strategic foresight to ensure the survival and revitalization of indigenous cultures and their profound knowledge systems.

\textbf{Cultural Erosion:} The processes of globalization and modernization indeed pose a threat to indigenous cultures, potentially marginalizing or causing the loss of their unique traditional knowledge and cultural practices \cite{un2019indigenous}. For a long time, indigenous cultures have not received sufficient attention within mainstream perspectives, even though they possess rich oral traditions and profound connections with nature.
Historical policies, such as those of residential schools for indigenous peoples, may force many to abandon their languages and cultures in favor of mainstream society. These assimilationist policies not only weaken the cultural identity of indigenous peoples but also put their traditional knowledge at risk of being lost \cite{colonization_impact}.

\textbf{Language Loss:} The disappearance of indigenous languages is a global issue that signifies more than just the loss of a means of communication; it represents the vanishing of entire cultures and knowledge systems \cite{cvf2e2rduqb6sudn8ju0,Chen2023EvaluatingSS}. The number of fluent speakers of indigenous languages is dwindling rapidly, with some languages having very few remaining speakers. Language is a vital carrier of traditional knowledge, and its loss can lead to the disappearance of associated cultural practices, spiritual beliefs, and historical memories.

\textbf{Intellectual Property Issues:} Indigenous knowledge is often commercialized or misused without proper authorization or compensation, infringing upon the rights of indigenous communities \cite{TonePahHote2022TheCO,Oyelude}. For instance, traditional healing methods may be appropriated by pharmaceutical companies and patented \cite{Orozco2013TheRO}, while indigenous songs may be adapted and copyrighted without any recognition or benefit shared with the originating communities. To address this, it is necessary to enforce intellectual property laws more rigorously, ensuring that indigenous communities receive appropriate acknowledgment and compensation when their traditional knowledge is utilized.

\textbf{Lack of Documentation:} Much of the traditional knowledge is passed down orally and lacks systematic recording and documentation, making it susceptible to being lost \cite{traditional2024,Vijayan2022}. To safeguard this knowledge, measures such as the creation of databases, registers, and registries must be taken to facilitate the preservation and transmission of traditional knowledge \cite{lekhi2024loss}. However, documentation should not be seen as an end in itself but as part of a broader intellectual property strategy that ensures the protection of the intellectual property rights of traditional knowledge while recording and disseminating it. This includes ensuring the free, prior, and informed consent of the holders of traditional knowledge and the implementation of fair and equitable benefit-sharing principles.

The intricate challenges of preserving indigenous knowledge are further compounded by the dynamic interplay between the needs of industry and science and the traditional practices of indigenous communities. As we have identified the key issues that hinder the preservation and appreciation of indigenous knowledge, it is now crucial to understand how these challenges intersect with the broader societal and economic contexts. The growing recognition of indigenous knowledge's value within industrial and academic spheres brings both opportunities and dilemmas. It is at this intersection where the true complexity of the situation becomes apparent, as the push for innovation and sustainability in modern industries must be balanced with the respect and protection of indigenous cultures and their knowledge systems. This balance is delicate and requires a nuanced approach that acknowledges the unique position of indigenous knowledge within the global intellectual property landscape, ensuring that it is not only preserved but also integrated in a way that honors its origins and significance.

In particular, the intersection of industrial and scientific demands with the preservation and dissemination of indigenous knowledge presents a complex landscape. On one side, there is a burgeoning recognition within industry and academia of the necessity to safeguard and propagate the rich tapestry of indigenous wisdom, which encompasses sustainable practices and innovative approaches that have been honed over generations. However, this sector often grapples with the challenge of integrating indigenous knowledge into existing intellectual property frameworks in a manner that respects its collective and cultural significance \cite{care_principles2023}. Conversely, from the indigenous perspective, there are inherent difficulties in maintaining the integrity and continuity of their knowledge systems amidst the pressures of globalization and cultural assimilation. The urgency and necessity of preserving indigenous knowledge are underscored by the very real threats of erosion and loss, which, if unaddressed, could result in the irretrievable disappearance of invaluable cultural heritage and practical wisdom \cite{science2022indigenous}.

The following discussion will first introduce a technological framework for a sustainable knowledge supply chain. It will then review the current technologies being used at each stage of the supply chain, identify gaps where technologies are lacking, and propose potential solutions. This approach aims to create a future that is both technologically advanced and culturally sensitive, aligning with the global dialogue on indigenous rights and sustainable knowledge practices.

This review article focuses on the crucial issue of indigenous knowledge dissemination and protection. Its main contributions are as follows:
\begin{itemize}
\item We comprehensively analyzed the challenges in indigenous knowledge transmission and protection. Based on in-depth research of existing technologies, we innovatively proposed a sustainable indigenous knowledge supply chain framework. This framework integrates knowledge protection and dissemination across the entire supply chain, enabling a holistic approach to empowering indigenous knowledge protection and dissemination.

\item In the transmission and protection process of the indigenous knowledge supply chain, we prioritized indigenous data sovereignty. By designing a reverse loop for knowledge value realization, we facilitated the flow of value back to indigenous communities. This supports indigenous peoples in maintaining their livelihoods locally, fostering a virtuous ecosystem for indigenous knowledge.

\item On the basis of the proposed indigenous knowledge supply chain, we meticulously examined the key advanced technologies in the critical stages of knowledge collection and dissemination. We analyzed existing technologies, identified gaps, and explored promising technologies. We particularly emphasized the potential and value of advanced technologies in enabling indigenous knowledge protection and transmission, with detailed discussions on the applicability of these technologies in specific indigenous knowledge collection and protection scenarios.

\item Building on the comprehensive explanation of the proposed indigenous knowledge supply chain framework, we elaborated on how our framework effectively addresses existing challenges in indigenous knowledge protection and transmission. We further provided promising solutions from a technical perspective to overcome these challenges.

\item Drawing on previous work and our proposed knowledge protection framework, we detailed promising future research directions for indigenous knowledge protection and transmission from a technical standpoint. These directions offer a promising guideline for future research, introducing new perspectives such as security, computational power, resilience, and sustainability.
\end{itemize}

This paper is structured as follows: Section 1 emphasizes the significance of indigenous knowledge for human civilization and social development, and highlights the critical topic of indigenous knowledge protection. Section 2 identifies the current challenges in indigenous knowledge protection and stresses the urgent need for a sustainable protection framework. Section 3 presents a sustainable indigenous knowledge protection framework, detailing the participants, roles, and key technologies in each component, and technically justifies the importance of these technologies for indigenous knowledge preservation. Section 4 explains how the proposed framework addresses existing challenges and elaborates on the application of promising technologies in indigenous knowledge supply chain development. Section 5 explores future research directions from the perspectives of security, resilience, and sustainability. Section 6 summarizes the entire paper.

\begin{figure*}[t]
\centering
\includegraphics[width=16.5cm]  {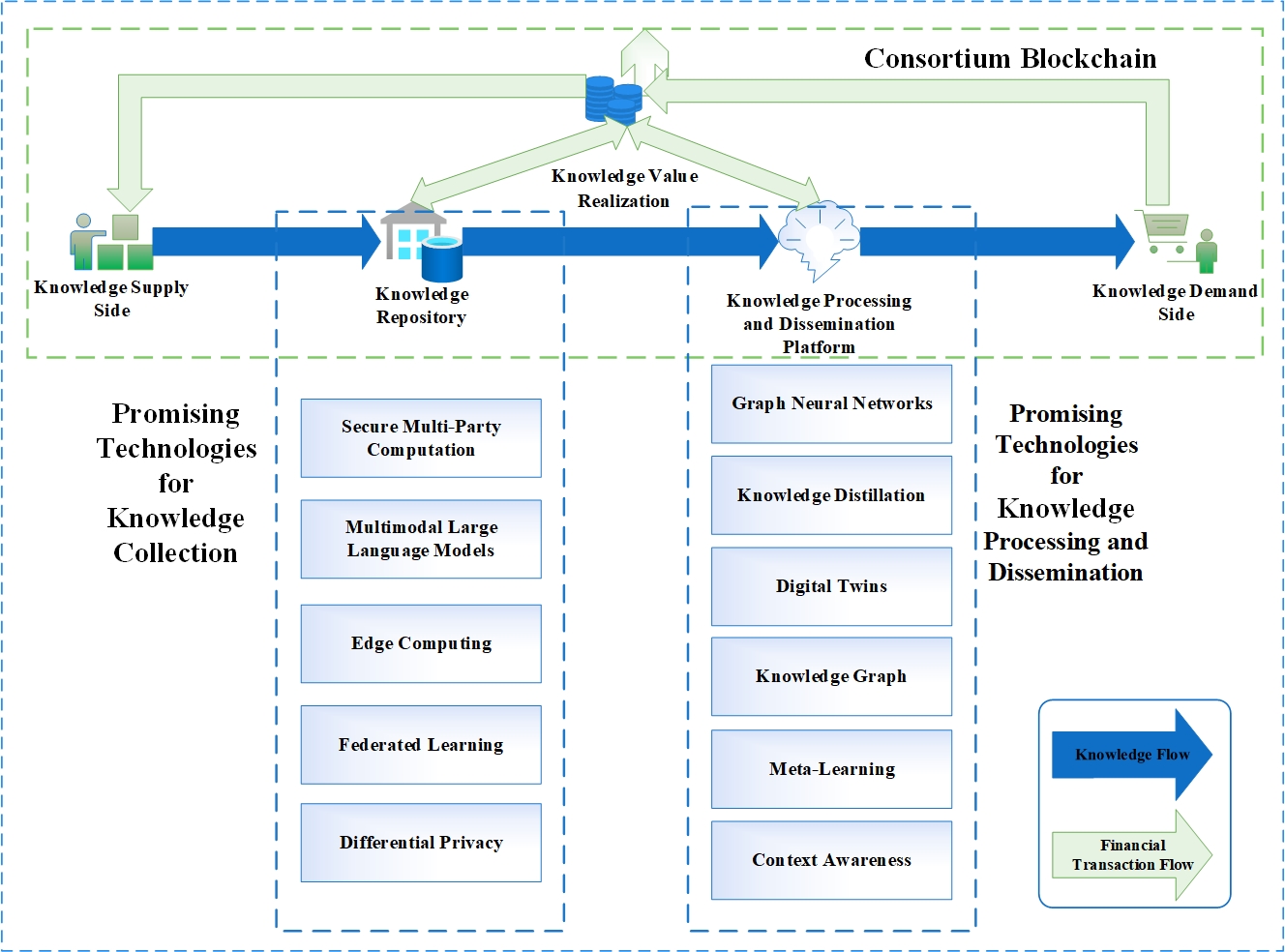}
\caption{\small Sustainable Trustworthy Indigenous Knowledge Supply Chain Framework, where the green dashed line box represents the entire consortium blockchain, the blue green dashed line box represents the key supporting technologies for the corresponding stage of the supply chain, and the solid line box at the bottom right is the legend.}
\label{F:multimodal}
\end{figure*}

\section{Knowledge Supply Chain for Preservation and Dissemination of Indigenous Knowledge}
This section introduces the sustainable and trustworthy indigenous knowledge supply chain framework, exploring the functions and interactions of its components and stakeholders. It emphasizes the significance of consortium blockchain in protecting indigenous knowledge sovereignty and data security, thereby facilitating the sustainable management of indigenous knowledge; see Figure \ref{F:multimodal}. Within this framework, we meticulously examine the existing technologies and identify the gaps and challenges in key components of the supply chain. By pinpointing these issues, we aim to highlight promising advanced technologies that can safeguard the indigenous knowledge supply chain from multiple perspectives, including sustainability, security, and efficiency.
\subsection{Sustainable and Trustworthy Indigenous Knowledge Supply Chain Structure}
\begin{table}[htbp]
	\centering
	\caption{Components, Stakeholders, and Function and Role  of the Sustainable Supply Chain of Indigenous Knowledge}
\begin{tabular}{@{}p{\colthreewidth}p{\colfourwidth}p{\colsixwidth}@{}}
		\hline
		\textbf{Component} & \textbf{Stakeholders} & \textbf{Function and Role} \\
		\hline
		Supply End & Indigenous communities & Provide the foundational content of indigenous knowledge, including traditional agricultural techniques, ecological conservation methods, and cultural practices. These are transmitted within communities through oral traditions and practical demonstrations, forming the raw material for subsequent collection and processing. \\
		\hline
		Knowledge Collection & Academic research institutions, NGOs, commercial enterprises & Transform indigenous knowledge from oral and practical forms into storable and shareable formats such as text, images, and videos through field investigations, interviews, and recordings. Stakeholders must possess flexible technical skills to ensure accurate collection and respect for indigenous intellectual property rights. \\
		\hline
		Knowledge Processing and Dissemination & Knowledge management platforms, educational institutions, media & Organize, classify, and store indigenous knowledge, and distribute it in formats suitable for different audiences through digital technologies, educational course design, and media dissemination. Examples include developing online courses, producing documentaries, and creating knowledge databases to enhance accessibility and practicality. \\
		\hline
		Demand End & Educational institutions, healthcare providers, environmental management departments, tourism industry & Have practical needs for indigenous knowledge, applying it in education, healthcare, environmental management, and tourism to promote heritage and economic development. \\
		\hline
		Knowledge Value Realization & Indigenous communities, research institutions, enterprises, entrepreneurs, investors & Negotiate supply chain pricing, guarantees, and settlements; commercialize and apply indigenous knowledge to generate economic benefits, applying it to scientific research, cultural development, product development. The key is to channel economic benefits back to indigenous communities and the other stakeholders to protect their interests and encourage their participation in knowledge preservation and transmission, ensuring the sustainability of the entire supply chain. \\
		\hline
	\end{tabular}
	\label{tab:overview}
\end{table}

The sustainable and trustworthy supply chain of indigenous knowledge encompasses five main components: supply end, knowledge collection, knowledge processing and dissemination, demand end, and knowledge value realization; see Table \ref{tab:overview}. Each component involves various stakeholders who play crucial roles in the preservation, transmission, and application of indigenous knowledge, shown as follows.  
\begin{itemize}
\item \textbf{Supply Side}  The supply end of the indigenous knowledge supply chain primarily involves indigenous communities. These groups are the primary custodians and disseminators of indigenous knowledge, which includes traditional agricultural techniques, ecological conservation methods, and cultural practices \cite{Vijayan2022}. This knowledge is transmitted within communities through oral traditions and practical demonstrations, forming the raw material for subsequent collection and processing.

\item \textbf{Knowledge Collection}
The knowledge collection phase involves organizations and enterprises collaborating with indigenous communities to gather and document indigenous knowledge. These collectors can be academic research institutions, non-governmental organizations, or commercial enterprises. They transform indigenous knowledge from oral and practical forms into storable and shareable formats such as text, images, and videos through field investigations, interviews, and recordings. This phase also includes the categorization of knowledge into areas such as disaster mitigation, environmental protection, livestock management, crop cultivation, and aquaculture. The stakeholders in this phase must possess cross-cultural communication skills to ensure accurate knowledge collection and respect for indigenous intellectual property rights.

\item \textbf{Knowledge Processing and Dissemination}
The knowledge processing and dissemination phase focuses on making indigenous knowledge more accessible and applicable. Key stakeholders in this phase include knowledge management platforms, educational institutions, and media organizations. They utilize digital technologies, educational course design, and media dissemination to process, organize, classify, and store indigenous knowledge, and distribute it in formats suitable for different audiences. Examples include developing online courses, producing documentaries, and creating knowledge databases, all of which enhance the accessibility and practicality of indigenous knowledge.
\item \textbf{Demand End}
The demand end encompasses various sectors such as education, science, healthcare, environmental management, and tourism, which have practical needs for indigenous knowledge. Educational institutions can incorporate indigenous knowledge into their curricula to foster students' cultural diversity and environmental protection awareness. Healthcare institutions can draw on indigenous traditional medical knowledge to develop new treatments. Environmental management departments can leverage indigenous ecological conservation experience to formulate more effective environmental policies. The tourism industry can develop tourism products themed around indigenous culture, promoting cultural heritage and economic development.
\item \textbf{Knowledge Value Realization}
The knowledge value realization phase focuses on commercializing and applying indigenous knowledge to generate economic benefits. Key stakeholders—indigenous peoples, research institutions, enterprises, entrepreneurs, and investors—collaborate to apply this knowledge in scientific research, cultural development, product design, service provision, and brand building, unlocking its scientific, cultural, and economic value. In particular, this phase ensures economic benefits flow back to indigenous communities, protecting their interests and incentivizing their role in knowledge preservation and transmission. To support this process, stakeholders often form an alliance-based affairs committee that negotiates supply chain pricing, guarantees, and settlements. Using consortium blockchain and smart contracts, the committee dynamically prices transactions and distributes benefits in real-time, ensuring all parties receive fair compensation. This approach promotes sustainability and equity across the entire supply chain ecosystem, balancing economic growth with cultural preservation.

\end{itemize}

\subsection{Consortium Blockchain Safeguarding Indigenous Knowledge Sovereignty and Data Security}
Data security and trustworthiness are paramount in achieving indigenous knowledge autonomy \cite{Berkes2022,MACKEY20222626}. Only with secure and trustworthy data can the value chains of each segment in the supply chain be fully connected and continuously increased, thereby fully protecting the intellectual property value of indigenous knowledge. Blockchain technology, particularly consortium blockchains \cite{Consortium1}, plays a vital role in this regard. The consortium blockchain operates by authorizing trusted organizations to collaboratively manage the network, ensuring data security and confidentiality. By sharing information appropriately among the alliance, it streamlines product flow and traceability, making it particularly suitable for supply chain scenarios. It can be applied across multiple segments of the indigenous knowledge sustainable supply chain, from the supply end to the knowledge value realization phase, ensuring the secure and transparent management of indigenous knowledge.
\begin{itemize}
\item \textbf{Safeguarding Supply Side} 
At the supply end, blockchain can be used to record and verify the origin and authenticity of indigenous knowledge. By creating a unique blockchain record for each knowledge module, the technology ensures that the source of the knowledge is trustworthy and cannot be tampered with during transmission. This not only helps protect the intellectual property rights of indigenous knowledge, ensuring its legality and authenticity, but also enables indigenous communities to better trace the flow of their knowledge through various stages of the supply chain. This improved traceability promotes sustainable practices by guaranteeing that the utilization and spread of indigenous knowledge are both transparent and accountable.
\item \textbf{Safeguarding Knowledge Collection}
During the knowledge collection phase, blockchain can be used to record the entire collection process and the contributions of each participant. This ensures the transparency and traceability of knowledge collection, establishing a fair mechanism that protects the rights and interests of indigenous communities.
\item \textbf{Safeguarding Knowledge Processing and Dissemination}
In the knowledge processing and dissemination phase, blockchain is generally used for managing access permissions and the dissemination process through smart contracts. This ensures that only authorized users can access specific knowledge modules and prevents the abuse and unauthorized dissemination of knowledge.
\item \textbf{Safeguarding Demand End}
 At the demand end, blockchain is mainly used for recording the usage of knowledge and user feedback. By storing this information on the blockchain, a feedback mechanism can be established to optimize knowledge processing and dissemination. This not only enhances the practicality of the knowledge and user satisfaction but also aligns with the principles of sustainable development. 
\item \textbf{Safeguarding Knowledge Value Realization}
In the knowledge value realization phase, blockchain could be utilized to manage commercial transactions and profit distribution through smart contracts. This ensures that commercial transactions are transparent and fair, and that profit distribution follows predetermined rules. This not only safeguards the economic interests of indigenous communities but also ensures the rational realization of knowledge value, thereby effectively propelling the sustainable development of the indigenous knowledge supply chain ecosystem.
\end{itemize}

\subsection{Advanced Technologies Facilitating Indigenous Knowledge Collection, Processing, and Dissemination}
In the earlier sections, we have explored the composition of the sustainable indigenous knowledge supply chain and highlighted how consortium blockchains can safeguard the supply chain from various perspectives, including security, sustainability, and efficiency. This subsection continues to focus on two crucial components of the supply chain: the knowledge collection and the knowledge processing and dissemination stages. We will review the technological advancements in these areas, identify existing challenges and gaps, and explore promising cutting-edge technologies that hold the potential to transform these stages.
\subsubsection{Knowledge Collection}
This section is divided into three main aspects, namely Solutions Developed, Gaps Remaining, and Promising Solutions, as detailed below:

\textbf{Solutions Developed:}

The following studies demonstrate how technical means have emerged as a crucial force in preserving indigenous cultures, offering a variety of approaches to maintain this knowledge, ranging from languages and traditional crafts to agricultural systems.

Addressing the challenge of low-resource languages, the authors introduce the development of Cendol, a collection of Indonesian Large Language Models (LLMs) designed to bridge the quality gap, particularly for Indonesian indigenous languages \cite{Cahyawijaya202414899}. These models are meticulously trained to capture the nuanced characteristics of these languages, thereby enhancing their effectiveness in tasks such as translation and text generation.

In another contribution, the authors put forward a personalized news media extraction and archival framework that focuses on traditional knowledge \cite{George2020463}. The system leverages ontology to assist journalists in identifying semantically matching news, with a specific emphasis on extracting and archiving content related to indigenous knowledge.

Exploring innovative applications, the researchers explore the use of Virtual Reality (VR) technology to preserve and share indigenous knowledge within communities \cite{Soosay2024379}. They also propose a project to create a virtual learning tool for sewing traditional ribbon skirts, utilizing a 360-degree camera. This approach aims to strengthen community bonds and provide connection points for dispersed community members.

Focusing on regions with high agro-biodiversity, the authors present a maximum entropy-based model to exploit the usage of Agricultural Heritage Systems (AHS) \cite{Bai2024670}. They further construct a map collecting potential AHS areas in China, validated through existing AHS. This method not only helps conserve traditional agricultural systems but also preserves the indigenous knowledge associated with them.

Beyond agriculture, the integration of AI and indigenous knowledge extends to environmental management and conservation. The authors introduce an Internet-of-Things (IoT) and AI-driven solution for mitigating human-wildlife conflict in agriculture \cite{Abed2025}. The developed approach integrates an ultrasonic sensor and a fine-tuned deep-learning model to identify and categorize animal species, triggering species-specific deterrents to protect crops.  This approach combines IoT technology with indigenous knowledge to enhance agricultural productivity and biodiversity conservation.

The following research explores how indigenous knowledge can provide new perspectives and solutions for modern technology fields, especially when dealing with global challenges like climate change and environmental issues. The combination of indigenous wisdom and modern technology reveals unique advantages. 

To address environmental conservation challenges, the authors discuss the integration of artificial intelligence (AI) in environmental conservation, particularly in the context of climate change \cite{Scoville202130}. They further highlight the role of AI in reshaping data collection, classification, and decision-making processes. The article touches on the ethical dilemmas and power dynamics that can arise when AI is used in conservation efforts, which are often lands managed by indigenous peoples.

In the context of climate change, the authors discuss the importance of indigenous knowledge in smart agriculture \cite{Ahire2023241}. Furthermore, they highlighted the need to combine traditional agricultural practices with modern technologies like wireless sensors and machine learning to create sustainable agricultural methods. This approach aims to mitigate the effects of climate change while preserving indigenous agricultural knowledge.

\textbf{Gaps Remaining:}

Despite advancements, significant technical limitations persist in the collection and preservation of indigenous knowledge. Current technologies often struggle to fully capture the nuances and complexities of indigenous languages and knowledge systems. In addition, the integration of indigenous knowledge with advanced technologies like AI requires significant computing power and specialized knowledge, which are not always accessible to indigenous communities. This can lead to dependency on external support and raise issues of data ownership and control. Moreover, ethical considerations and potential biases in AI systems pose risks to indigenous knowledge. To address these challenges, more advanced technologies are needed to enhance the accuracy and effectiveness of capturing indigenous knowledge, while also ensuring that these technologies are accessible and empower indigenous communities to independently manage their knowledge.

\textbf{Promising Solutions:}

\textbf{Secure Multi-Party Computation}
Secure Multi-Party Computation (SMPC) could enable multiple participants to collaboratively   compute the result of a function while keeping their individual inputs private from one another.
This technology ensures that each party's data remains confidential throughout the computation process, as no party discloses their input to others. Instead, they collaboratively compute the desired function and obtain the result without revealing any sensitive information \cite{SMPC1,SMPC2}. SMPC can be applied to the processing of indigenous knowledge to enhance its security and privacy. For example, when multiple indigenous communities want to jointly analyze their traditional ecological knowledge to develop a conservation plan, they can use SMPC to compute the desired results without disclosing their specific knowledge to each other. This helps protect the confidentiality of their traditional knowledge while enabling collaborative decision-making. SMPC provides a secure environment for the computation, preventing unauthorized access and leakage of the knowledge. Leveraging SMPC, indigenous peoples can actively participate in and monitor the entire knowledge production and training process, thereby ensuring greater control and oversight over how their knowledge is utilized.

\begin{figure*}[t]
	\centering
	\includegraphics[width=10.5cm]  {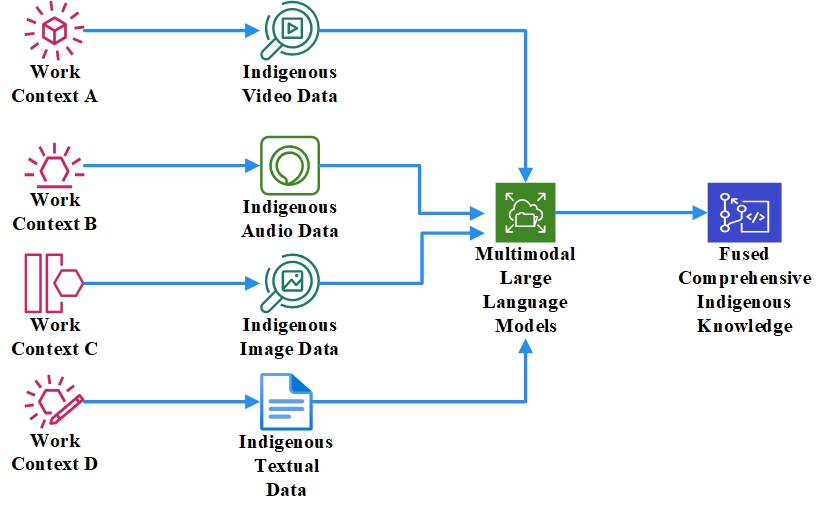}
	\caption{\small The Multimodal Large Language Models technology for fusing collected multimodal indigenous data, where the data from each modality is collected within the different indigenous work scenarios, with the authorization of the indigenous people. It is worth noting that the data collected in each work context is practically multimodal; here, the graph is only used for concise representation, with the main modality displayed. }
	\label{F:LLM}
\end{figure*}

\textbf{Multimodal Large Language Model} Multimodal Large Language Models (MLLMs) have the ability to handle and comprehend information from diverse sources, including text, images, videos, and audio  \cite{huang2024survey}; see Figure \ref{F:LLM}. For indigenous knowledge, this capability enables the simultaneous collection of stories narrated by elders (audio), related art pieces (images), and traditional problem solving solutions (video), which can then be transformed into digital multimodal representations by the model \cite{MLLM1,MLLM2,MLLM3,MLLM4,MLLM5}. This comprehensive approach allows for a richer documentation of cultural knowledge, preserving not only the content but also the context in which it is presented.
MLLMs can be trained to comprehend the relationship between language and visual content  \cite{bai2023qwen}. For instance, given an image of traditional attire, the MLLM could generate text that describes the garment, or conversely, provided with a text about crafting a traditional artifact, the model could produce an image that reflects the description. This dual capability facilitates a deeper understanding of cultural expressions by connecting verbal narratives with visual representations, thus providing a more immersive and interactive experience of indigenous cultures \cite{xiao2024comprehensive}.

\textbf{Edge Computing} In edge computing, services like computation, storage, and networking are relocated closer to the data source or the users, forming a distributed computing framework, thereby reducing latency and bandwidth usage \cite{EC1,EC2}. In the context of indigenous knowledge processing, edge computing can offer several advantages. For instance, it can enable real-time processing and analysis of indigenous knowledge data, allowing for faster decision-making and more timely interventions. This is particularly useful in scenarios where indigenous communities need to make immediate decisions based on their traditional knowledge, such as environmental monitoring or resource management. Edge computing can also help to preserve the privacy and security of indigenous knowledge by keeping data processing local and reducing the need to transmit sensitive information over long distances. In addition, edge computing can support the development of more personalized and context-aware applications for indigenous communities, as it allows for the integration of local knowledge and practices into the computing process. Overall, edge computing has the potential to enhance the processing and application of indigenous knowledge in a variety of contexts, from education and healthcare to environmental conservation and cultural heritage preservation.

\begin{figure*}[t]
	\centering
	\includegraphics[width=10.5cm]  {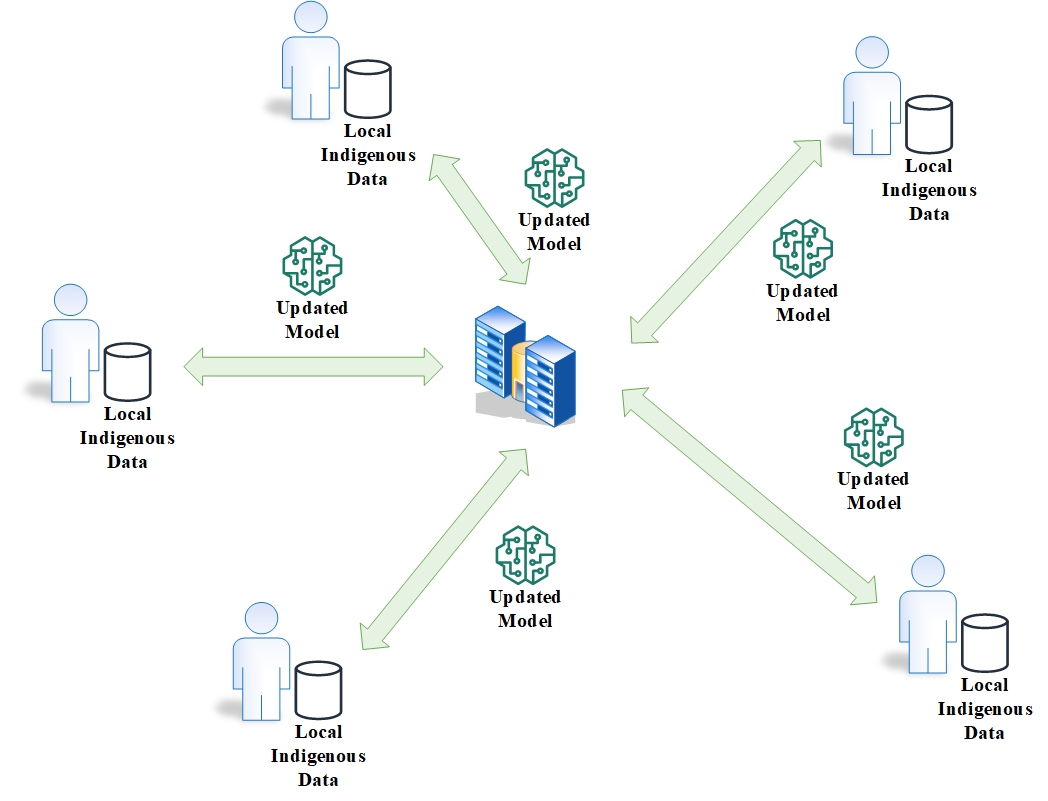}
	\caption{\small The Federated learning technology for joint multi-community indigenous knowledge model development, where each participating community only needs to interactively update its own model, while keeping the data local. 
		.}
	\label{F:Federated}
\end{figure*}

\textbf{Federated Learning} 
Federated learning is a machine learning methodology enabling numerous clients to collectively develop a model, with their data remaining local, orchestrated by a central server \cite{FL1,FL2,yuan2023Amplitude,Nan2024WWW}; see Figure \ref{F:Federated}. This approach allows for the processing of indigenous knowledge in a way that preserves its privacy and security. For example, different indigenous communities can participate in a federated learning framework to develop a model that captures their traditional knowledge without sharing their sensitive data with external parties. Each community can train a local model on it own data, and the central server can aggregate the model updates to improve the global model. This method not only protects the confidentiality of indigenous knowledge but also promotes collaboration and knowledge sharing among different communities \cite{FL1,Li2025A,Hu2024OFDMA,10948161}. In addition, federated learning can help to address the issue of data scarcity in some indigenous communities by leveraging the collective knowledge from multiple sources. Overall, federated learning provides a promising approach to the processing of indigenous knowledge, enabling more effective and inclusive knowledge collection and application.

\begin{figure*}[t]
	\centering
	\includegraphics[width=10.5cm]  {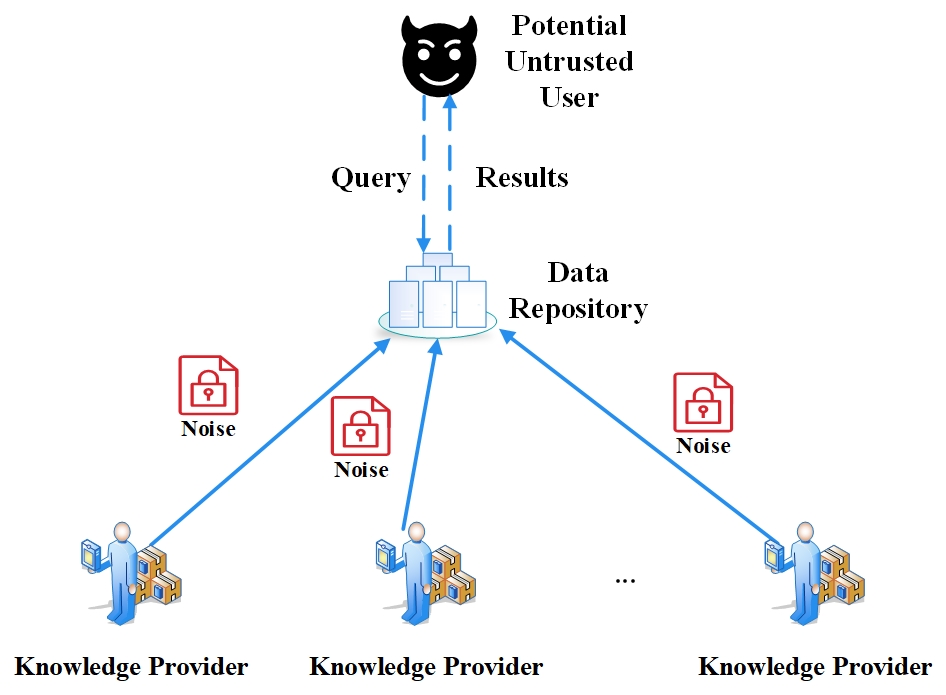}
	\caption{\small The Differential Privacy technology for joint multi-community indigenous data collection, where each user's raw data is perturbed with noise and encrypted into private data, ensuring that potential untrusted users cannot identify sensitive individual details. 
		.}
	\label{F:DP}
\end{figure*}
\textbf{Differential Privacy}
Differential Privacy is a privacy protection technique that safeguards individual information by adding noise to data or query results \cite{DP1,DP2,Shan2023Preserving,You2024Novel}; see Figure \ref{F:DP}. During indigenous knowledge collection, it shields privacy and data sovereignty, ensuring statistical analyses don't reveal sensitive individual details, thus preserving cultural integrity and privacy.
This technique offers a quantifiable method to measure and control privacy levels. The gathered indigenous data can be tailored to the sensitivity of the dataset by adjusting parameters, enabling more precise privacy protection. This is crucial for indigenous knowledge protection, as different knowledge types have varying sensitivities and cultural values.
In practice, differential privacy can be implemented in various ways, such as adding noise during data collection or processing.

\subsubsection{Knowledge Processing and Dissemination}
Consistent with the discussion on knowledge collection, this section is also divided into three parts for better elaboration, namely Solutions Developed, Gaps Remaining, and Promising Solutions, as detailed below:

\textbf{Solutions Developed:}

The integration of indigenous knowledge with modern technologies such as AI is a key focus in addressing various challenges, particularly in the context of climate change and sustainable development. For instance, the authors propose a unified approach that combines scientific and indigenous knowledge to create an early-warning system for assessing the local impact of climate change \cite{El-Beltagy202426}. This approach leverages indigenous knowledge to inform decision-making processes and improve crop management in agro-ecosystems affected by climate change. 

Similarly,  the authors present a model that integrates indigenous knowledge, climate data, and vegetation index to predict favorable weather seasons for crop cultivation, crop monitoring, and crop health prediction \cite{Nyetanyane20203}. The model utilizes machine learning to provide small-scale farmers with more accurate and location-specific weather information, addressing the limitations of traditional indigenous knowledge in the face of climate change and deforestation.

In another contribution, the authors  discuss the use of holistic systems thinking and applied geospatial ethics to address the challenges of the UN Sustainable Development Goals (SDGs) \cite{Caudill2024}. They further highlight the importance of integrating traditional and indigenous knowledge systems with geospatial technologies to create more inclusive and holistic approaches to sustainable development.

In the realm of agriculture, several studies focus on how AI and indigenous knowledge can be combined to support decision-making and enhance productivity. The researchers introduce an intelligent agricultural knowledge management system designed to assist farmers and agricultural extension workers in developing nations \cite{Buitendag2024}. 
The proposed approach uses knowledge objects to capture indigenous knowledge and enhances knowledge discovery through metadata and annotations. This addresses challenges in knowledge operations and decision-making for farmers and agricultural extension workers.

Targeting at selecting appropriate crops for small-scale farmers, the authors report to have achieved 99\% accuracy through the collection of climatic and soil-related data and multi-category classifying techniques \cite{Thothela2023}. The resulting framework utilizes machine learning and collected indigenous knowledge to integrate soil and climate factors, supporting small-scale farmers in decision-making processes.

In addition, emphasizing the importance of integrating local and indigenous knowledge with scientific knowledge, the scholars provide an overview of global control efforts to combat desertification \cite{Ahmed2024}. They argue that capacity building, education, and training should be more comprehensive for local communities, and that sustainable land management practices and advanced technologies can help standardize procedures and accurately assess the extent of desertification.

Focusing on the indigenous knowledge-based enables decision-making management in Australia's Kakadu National Park, the authors describe a co-production mechanism that weaves indigenous knowledge, artificial intelligence, and technical data  \cite{Robinson2022}. Furthermore, they highlight the importance of indigenous-led governance of research activities and the integration of qualitative indigenous assessments with quantitative ecological data. 

To better develop an intelligent system for comprehensive pollution monitoring, the authors discuss the integration of industrial technologies and indigenous knowledge \cite{Ramba2023}. In addition, they argue that machine learning can be applied to automate decision-making processes, thereby improving the system's ability to forecast outcomes precisely. They emphasize the need for indigenous knowledge to be a core partners in this process, both as contributors of traditional knowledge and priority communities for capacity development.

Recognizing the importance of the ethical and cultural considerations in integrating AI with indigenous knowledge, the authors examine how traditionally territorializing technologies can be paired with indigenous knowledge and protocols to operate between signal and noise, rendering perverse changes in the landscape comprehensible \cite{Charbel2020708}. They further present a methodology defined as 'decoding' and 'recoding', using four distinct case studies in the Arctic to address various scales and targets.

For better predicting the antiplasmodial potential of plant species, a unique dataset is introduced concerning the antiplasmodial activity exhibited by distinct flowering plant families, further showcasing the efficacy of AI \cite{Richard-Bollans2023}. In addition, they compare the predictive capability of various algorithms and ethnobotanical selection approaches, showing that machine learning models have higher precision. 
They argue that AI combined with indigenous knowledge has the potential to speed up the process of discovering new compounds derived from plants.

In another contribution, drawing inspiration from the five criteria devised by a Māori specialist, the authors explore 
 controversial matters, formulating a new indigenous perspective on AI \cite{Munn20241673}. Furthermore, they explore applying each test to data-driven systems with examples, challenging current AI priorities, and offering a framework for reconsidering AI based on indigenous knowledge.

\textbf{Gaps Remaining:}

The current state of knowledge processing, handling, and dissemination technologies in the context of indigenous knowledge is marked by a growing recognition of the value of integrating traditional and modern approaches. However, several technical limitations and areas for improvement remain. 
First of all, its deep contextual and holistic nature, intertwined with cultural, ecological, and spiritual practices, means that extracting and isolating specific elements for integration into modern technological frameworks risks losing their original meaning and coherence. In addition, the cultural mismatch between modern technologies (such as AI) developed within contemporary scientific paradigms and the cultural values and cognitive frameworks of indigenous communities can lead to disconnections and reduced effectiveness in integration efforts. Furthermore, the scarcity and quality of data, compounded by potential biases and lack of structure, pose significant hurdles in effectively integrating indigenous knowledge with modern technologies like AI.

\textbf{Promising Solutions:}

\textbf{Graph Neural Network}
Graph Neural Network (GNN) is a specialized neural network architecture tailored for handling data that is organized in a graph format, enabling the modeling of complex relationships between entities~\cite{GNN1,GNN2,GNN3,10518175}. In the context of processing and distributing indigenous knowledge, GNNs can serve as a powerful tool for analysis. They can be used to analyze graph-structured data, where indigenous knowledge modules and users are represented as nodes. Through this approach, GNNs can perform tasks such as the analysis of knowledge sharing and usage patterns, the identification of key indigenous knowledge modules relevant to specific user groups, and the optimization of indigenous knowledge distribution, thereby improving accessibility and practicality, promoting preservation and transmission of indigenous knowledge.

\begin{figure*}[t]
	\centering
	\includegraphics[width=8.5cm]  {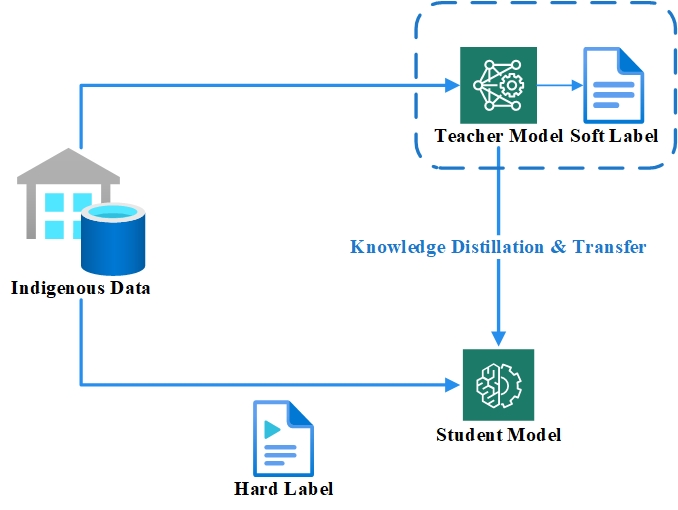}
	\caption{\small The Knowledge Distillation technology is a model compression technique transferring knowledge from a complex teacher model to a simpler student model, where the essence lies in training with the teacher model's soft labels and ground-truth hard labels to maintain high performance while achieving a smaller size and faster inference speed.}
	\label{F:KD}
\end{figure*}

\textbf{Knowledge Distillation}
Knowledge Distillation is a method for compressing models by extracting the knowledge from a sophisticated, high-performing teacher model and embedding it into a more streamlined and efficient student model. This process allows the student model to deliver comparable performance with a smaller model size and faster inference speed \cite{KD1,KD2}; see Figure \ref{F:KD}. This technology trains the student model using the output probabilities (soft labels) of the teacher model and true labels (hard labels), helping the student model capture the teacher model's decision-making process and implicit knowledge. Knowledge Distillation technology simplifies and optimizes indigenous knowledge, significantly enhancing its accessibility and practicality. This technology extracts and transforms the core elements of indigenous knowledge into an applicable, light-weight form.  Moreover, through knowledge distillation technology, the model constructed from the indigenous people's data can be transferred to a smaller-scale model, thereby effectively reducing the risk of leakage of the original data of the indigenous people \cite{KD1,KD2}. Meanwhile, the smaller model size also potentially reduces the computational power and deployment costs.

\begin{figure*}[t]
	\centering
	\includegraphics[width=13.5cm]  {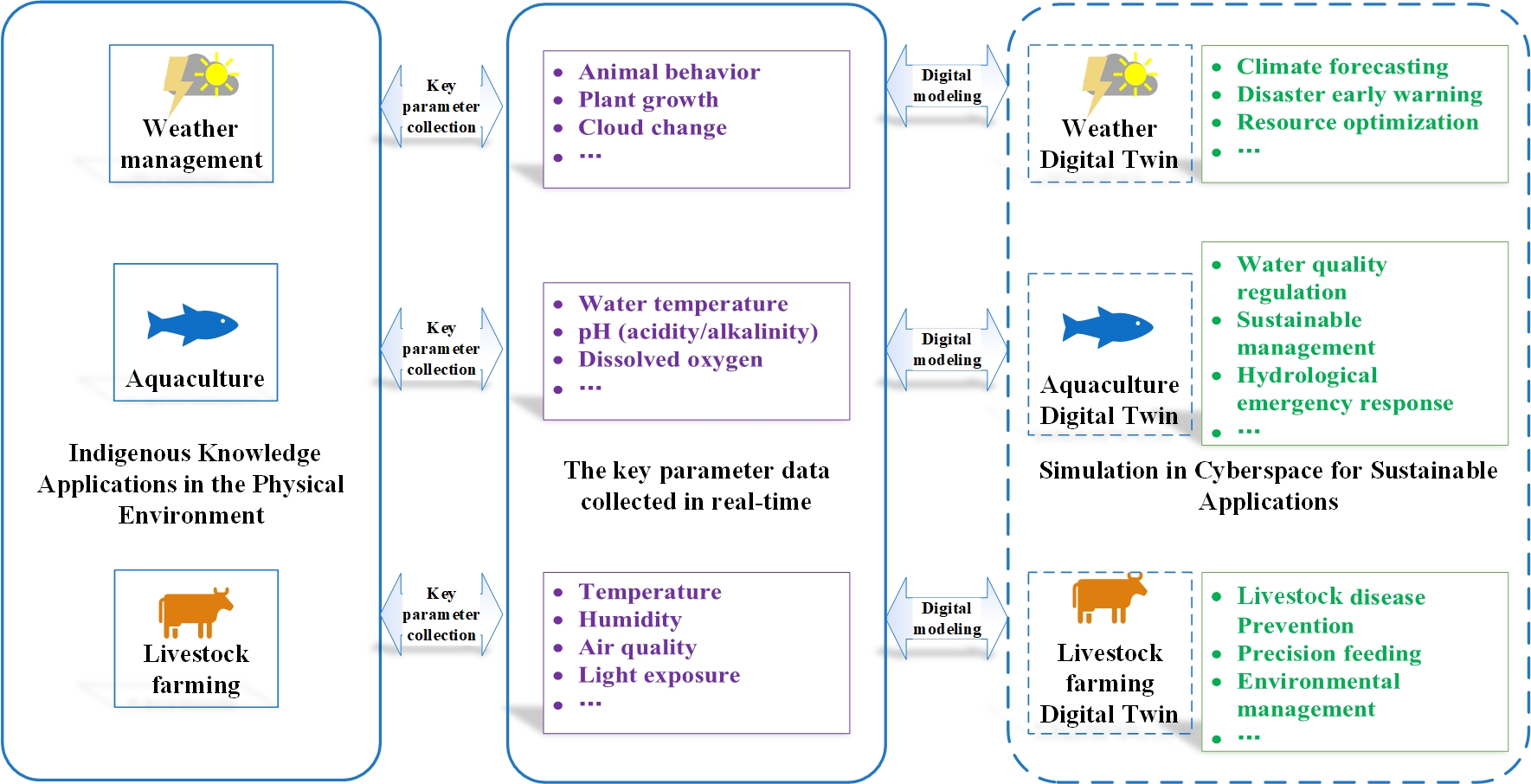}
	\caption{\small The Digital Twins technology is an interactive digital modeling technique by creating digital replicas from the original physical environment, where typical scenarios displayed as weather perception and agriculture, are used for highlighting potential applications. }
	\label{F:DT}
\end{figure*}

\textbf{Digital Twin}  By creating a digital replica of the indigenous knowledge modules, digital twin technology enables a more efficient and interactive way to manage and share indigenous knowledge; see Figure \ref{F:DT}. For instance, the digital twin can simulate the interactions between users and knowledge modules, providing a more immersive and engaging experience for learners \cite{DT1,DT2}. This technology can also facilitate the preservation and transmission of indigenous knowledge by allowing for real-time updates and modifications to the digital twin, ensuring that the knowledge remains relevant and accessible. In addition, digital twins can support the integration of indigenous knowledge into various applications by providing a versatile and adaptable platform for knowledge dissemination. Leveraging digital twin technology, it is possible to simulate and reproduce complex contexts, thereby preserving the dynamic and holistic nature of indigenous knowledge. This approach helps prevent the loss of its core significance and value when integrating indigenous knowledge into modern technological frameworks.

\textbf{Knowledge Graph}
Knowledge graphs built using graph databases create structured knowledge systems by linking fragmented knowledge elements. This technology enables in-depth analysis and integration of knowledge.  By building a knowledge graph, indigenous knowledge entities, application scenario entities, user entities, and their attributes can be represented as nodes. The relationships between these entities are represented as edges \cite{KG2}. This allows for a more organized and interconnected representation of indigenous knowledge, facilitating its processing and dissemination. the knowledge graph, by means of GNNs, can be used to analyze the relationships between different knowledge modules, identify key knowledge areas, and recommend relevant knowledge to users based on their interests and needs \cite{su14105881}, thereby helping to avoid mismatches between the contemporary scientific paradigms and the cognitive frameworks of indigenous communities \cite{KG1}.

\begin{figure*}[t]
	\centering
	\includegraphics[width=10.5cm]  {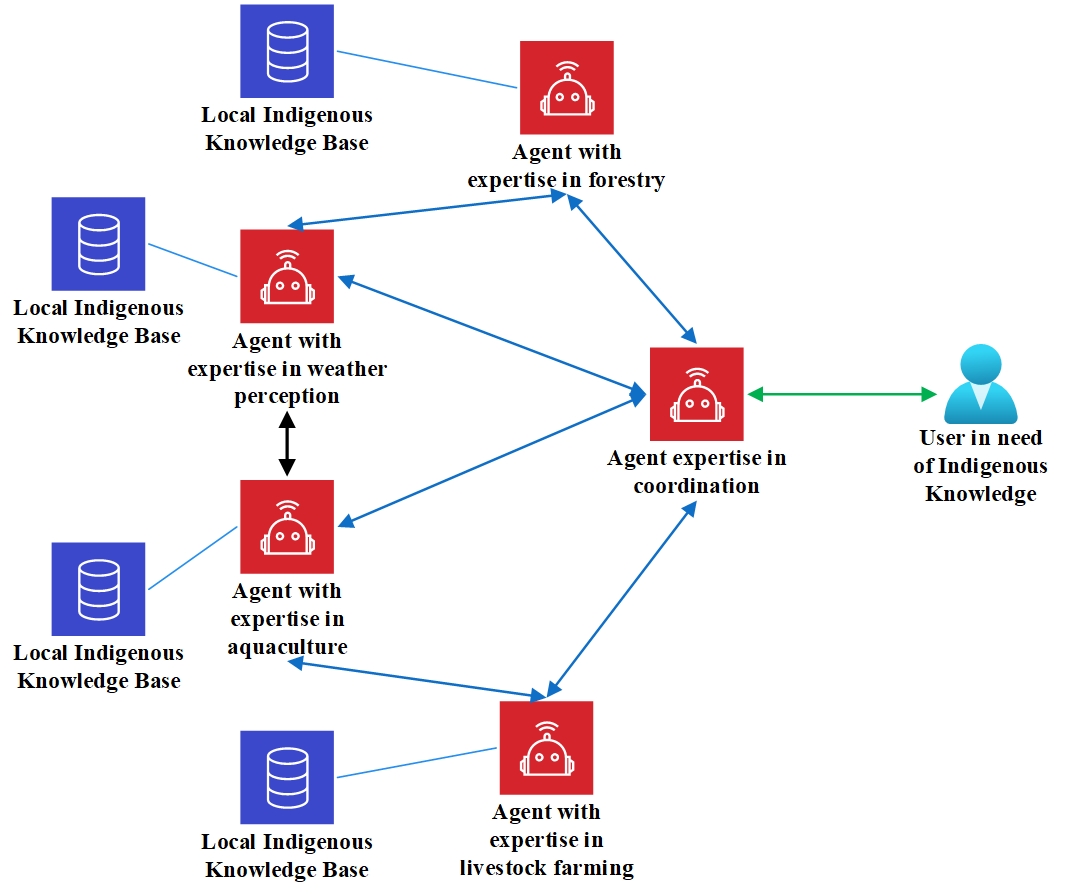}
	\caption{\small The Multi-Agent Systems technology enables effective collaboration among agents with diverse domain knowledge and skills, facilitating the interaction between indigenous knowledge and knowledge seekers in the downstream supply chain.}
	\label{F:MAS}
\end{figure*}

\textbf{Multi-Agent System}
 Multi-Agent Systems (MASs) enable effective collaboration among agents with diverse domain knowledge and skills, thereby accelerating the connection and recommendation between indigenous knowledge and knowledge seekers in the downstream supply chain \cite{MAS1,MAS2,You2024Privacy}; see Figure \ref{F:MAS}. For example, different knowledge modules such as traditional practices, cultural stories, and ecological wisdom can be represented as agents, and these agents can interact and collaborate to process and distribute indigenous knowledge. Specifically, MAS can facilitate collaborative processing of indigenous knowledge by enabling different agents to work together to process and analyze knowledge modules. In particular, the knowledge-demanding party first communicates with the coordinating agent. Subsequently, the coordinating agent communicates with the relevant agent who specializes in a particular indigenous field to obtain feedback. Based on this feedback, a comprehensive evaluation is conducted, and the appropriate information is then relayed back to the user in accordance with their intent. In short, MAS can enhance the accessibility and practicality of indigenous knowledge by enabling the knowledge demand side to interact with agents to access and utilize the knowledge.

\textbf{Meta-Learning}
Meta-learning, with its ability to "learn how to learn," offers efficient solutions for the preservation and processing of indigenous knowledge. Meta-learning enables efficient preservation and processing of indigenous knowledge by facilitating quick adaptation to new tasks and generalization with limited data \cite{Meta1,Meta2}. It enhances integration of fragmented knowledge through dynamic learning strategies in knowledge graph construction and multi-agent systems, and boosts adaptability and personalization of intelligent Q\&A systems and VR/Augmented Reality (AR) technologies. This increases processing efficiency, accelerates multi-scenario product applications, and opens new technological pathways for protecting, disseminating, and realizing the value of indigenous knowledge, including accelerating its value monetization. In addition, this approach allows AI systems to adapt and generalize better from smaller datasets, making it particularly useful in specific scenarios where data is scarce or of variable quality. Therefore, meta-learning could enhance the model's ability to quickly learn and make accurate predictions with minimal data, thus providing a promising solution for integrating indigenous knowledge into AI frameworks where data limitations are a significant concern.

\textbf{Context Awareness}
Context Awareness, which refers to systems that can perceive and respond to the context of users, such as their location, time, and preferences, can be effectively applied to the processing and dissemination of indigenous knowledge. By incorporating context-aware computing, the processing and dissemination of indigenous knowledge can be more personalized and relevant to the specific needs and situations of users \cite{CA3,CA2,CA1}. For example, context-aware systems can provide indigenous knowledge that is relevant to a user's current location or activity, enhancing the accessibility and practicality of the knowledge. This approach can also help to preserve and transmit indigenous knowledge in a more engaging and meaningful way, further assisting in addressing the issue of preserving the dynamic and holistic nature of indigenous knowledge when integrating it with specific application scenario.

\section{Discussion on Indigenous Knowledge Supply Chains: Protection, Preservation, and Future Technical Directions }
This examination of the knowledge supply chain structure for indigenous cognition underscores the pivotal role of sustainable practices in preserving cultural heritage and knowledge diversity. The subsequent sections will explore the multifaceted benefits of a supply chain approach that is inherently designed to protect and promote indigenous knowledge. We will discuss how such a structure can serve as a bulwark against cultural erosion by providing a stable platform for cultural practices to thrive; act as a safeguard for language preservation by creating avenues for language use and revitalization; address intellectual property concerns by establishing clear pathways for knowledge attribution and compensation; and streamline the documentation process to ensure the longevity of indigenous knowledge. By focusing on the supply chain's inherent capabilities, these sections will highlight how a thoughtfully constructed system can support indigenous communities in their quest for cultural continuity and knowledge sovereignty; see Table \ref{tab:rationale}.

\begin{table}[h!]
\centering
\begin{tabular}{{@{}p{\colonewidth}p{\coltwowidth}p{\colsixwidth}p{\colsixwidth}@{}}}
\hline
\textbf{Challenge} & \textbf{Current Challenge} & \textbf{Knowledge Supply Chain Solution} & \textbf{Rationale for Knowledge Supply Chain} \\
\hline
\textbf{Cultural Erosion} & The loss of cultural identity and traditional practices due to assimilation and dominance of external cultures. & The supply chain integrates advanced technologies and promotes respectful dissemination of cultural products. & Mitigates erosion by ensuring cultural products are valued for their authenticity, fostering an environment that nurtures cultural preservation. \\
\hline
\textbf{Language Loss} & The disappearance of indigenous languages leads to a loss of cultural heritage and diversity. & The supply chain supports access to necessary resources and advanced technologies, enabling the development of digital tools and online courses. & Preserves languages by making them accessible for learning and use, thus promoting cultural continuity and diversity. \\
\hline
\textbf{Intellectual Property Issues} & Indigenous communities often lack the means to protect their intellectual property from unauthorized use. & The supply chain integrates blockchain technology for enhanced tracking and employs smart contracts to facilitate automated compensation. In addition, it utilizes federated learning and SMPC to foster a decentralized and privacy-preserving approach to data management and analysis. & Protects indigenous knowledge by ensuring transparent and fair use, empowering communities with control over their intellectual property. \\
\hline
\textbf{Lack of Documentation} & Inadequate documentation of indigenous knowledge risks the loss of valuable cultural and practical wisdom. & The Supply Chain with technological innovations like blockchain, knowledge
graphs, and LLM presents a comprehensive strategy for the protection and promotion of indigenous knowledge. & Ensures that knowledge is systematically recorded and stored, providing a foundation for education, research, and cultural inheritance. \\
\hline
\end{tabular}

\caption{Detailed Challenges and Solutions within the knowledge supply chain for Indigenous Cognition}
\label{tab:rationale}
\end{table}

\subsection{Combating Cultural Erosion and Future Technical Directions}
Cultural erosion refers to the phenomenon where a culture gradually loses its uniqueness and value during contact with another culture, typically occurring in the process of influence and assimilation by a dominant culture over a weaker one. To address this challenge, protecting and promoting the culture of indigenous peoples becomes particularly important.

In terms of indigenous knowledge supply chain management, ensuring that the dissemination of cultural products and knowledge respects their original context and value is key to increasing the emphasis on cultural protection. This includes establishing a reasonable indigenous knowledge supply chain structure and strengthening long-term cooperative relationships with suppliers and partners.

Furthermore, advanced technologies like LLM, GNN, knowledge graph, digital twins, and MAS can play a significant role in this process. GNN and knowledge graph technology can assist in analyzing and predicting the market demand for cultural products, thereby providing data support for the protection and promotion of indigenous cultures. On the other hand, LLM and knowledge graphs can be utilized to analyze consumer behavior, gaining a better understanding of the interests and needs regarding indigenous cultural products. Moreover, digital twins can be employed to create virtual museums or exhibitions, allowing a global audience to experience and learn about indigenous cultures, thereby enhancing their visibility and impact.

As such, through trustworthy blockchain integrated with advanced technologies in indigenous knowledge supply chain management, cultural erosion can be effectively addressed, and the culture of indigenous peoples can be protected and promoted, ensuring that cultural heritage is maintained and passed down from generation to generation. This not only helps to protect cultural diversity but also promotes cultural innovation and sustainable development.

\subsection{Mitigating Language Loss and Future Technical Directions}

Language loss is a global issue, signifying not just the disappearance of a means of communication but also the loss of rich cultural and historical knowledge. Knowledge supply chain management plays a crucial role in addressing this challenge by supporting language preservation projects, which can effectively promote the documentation, conservation, and revival of endangered languages. The involvement of knowledge supply chain management ensures that language preservation projects have access to necessary resources and technical support, enabling the development of digital tools and online courses. These tools and courses not only provide a convenient learning path for language learners but also offer a platform for communication and collaboration among language inheritors and researchers.

 Through these advanced technologies like differential privacy, edge computing, SMPC, knowledge graph, digital twins, context awareness, and MLLMs, users of endangered languages can document invaluable indigenous knowledge, leaving behind valuable materials for future research and learning. At the same time, these technologies make the learning and dissemination of languages no longer limited by geography and time, allowing people from all over the world to access and learn these languages. Online course products based on these advanced technologies can also incorporate multimedia elements, such as videos, audio, and interactive exercises, making the learning process more vivid and effective. Moreover, these products can serve as carriers of cultural heritage, enhancing people's understanding and interest in the culture behind these languages by showcasing cultural customs, arts, and history related to the languages.

 Furthermore, through the value realization component of the sustainable indigenous knowledge supply chain, more indigenous people are able to monetize their knowledge. This reduces the need for them to migrate for work, providing them with greater options, such as staying in their local communities to better support and promote the development and transmission of their local language culture.

\subsection{Securing Intellectual Property and Future Technical Directions}

The protection of indigenous knowledge within the global intellectual property landscape is a critical challenge that knowledge supply chain management can help address. By implementing clear legal frameworks and cooperation agreements, we can ensure that indigenous communities receive proper authorization and compensation for the use of their knowledge. This not only involves legal recognition of their rights but also provides them with the tools to manage their intellectual property.

 Blockchain, federated learning, differential privacy, and SMPC offer innovative solutions to safeguard indigenous knowledge. For instance, blockchain can be used to monitor and track the use of this knowledge, preventing unauthorized exploitation. Blockchain's immutability ensures secure and transparent record-keeping for intellectual property, while federated learning, differential privacy, and SMPC contribute to a decentralized and privacy-preserving approach to data management and analysis.

Integrating these technologies can enhance the protection of indigenous knowledge, ensuring its rational use and dissemination while respecting the rights of indigenous communities. Smart contracts, powered by blockchain and AI, can automate intellectual property licensing and royalty payments, securing the economic benefits realization for indigenous peoples from the commercial use of their knowledge.

\subsection{ Facilitating Indigenous Knowledge Documentation and Future Technical Directions}

Knowledge supply chain management's emphasis on transparency and traceability is essential for the preservation of indigenous knowledge. This approach ensures that the rich systems of traditional knowledge held by indigenous peoples, encompassing their medicinal practices, agricultural techniques, and artisanal skills, are not only remembered but also passed down to future generations. The application of knowledge supply chain management principles allows for the systematic documentation and storage of this knowledge, which is often intricately linked to the cultural heritage and lifestyle of indigenous communities. By integrating methods from knowledge supply chain management, we can safeguard both cultural and biological diversity, recognizing the interconnectedness of indigenous knowledge and the natural world.

Community-led sustainable livelihood projects are a prime example of how knowledge supply chain management can empower indigenous communities. These projects enable communities to actively participate in the stewardship of natural resources, which is fundamental to their cultural heritage, the well-being of their members, and the economic health of their regions. The integration of technologies such as blockchain, which provides a secure and transparent record-keeping system, can significantly enhance these efforts. Blockchain's decentralized and tamper-proof nature ensures that data related to indigenous knowledge is maintained with integrity, offering a reliable platform for managing intellectual property rights and transactions in a way that benefits the communities.

Furthermore, the use of meta-learning, knowledge graphs, and LLMs can greatly assist in the preservation and dissemination of indigenous knowledge. Knowledge graphs offer a visual representation of complex information, making the cultural and practical knowledge of indigenous peoples more accessible and understandable. They provide a structured way to organize and retrieve information, which is vital for education and research purposes. Meanwhile, LLMs can be employed to translate and interpret indigenous languages, preserving linguistic diversity and facilitating communication within and outside the communities. Meta-learning enables efficient preservation and dissemination of indigenous knowledge by integrating fragmented data, enhancing adaptability in knowledge graphs, and accelerating multi-scenario applications and value realization. These technologies, when used responsibly, can enhance the self-determination of indigenous peoples and support their efforts to maintain their unique cultures and environments for generations to come.

As such, the integration of knowledge supply chain management with technological innovations like blockchain, knowledge graphs, and LLMS presents a comprehensive strategy for the protection and promotion of indigenous knowledge. This strategy not only ensures the preservation of cultural diversity but also contributes to the broader goal of global biodiversity conservation. By providing indigenous communities with the tools and platforms to manage their knowledge resources effectively, we can support their autonomy and promote sustainable development that honors and respects their traditions and wisdom.

\section{Open Issues for Future Refinement}

\subsection{Identifying the Authenticity of Indigenous Knowledge}
The sustainable supply chain for indigenous knowledge can boost Intellectual Property (IP) protection and human civilization. But it may also encourage forgeries for profit \cite{Ahmad2024123}. Such forgeries are cultural theft, often disguised as "authoritative" information to meet market demands and make money, especially in the era of Artificial Intelligence Generated Content (AIGC), which harms the environment and worsens social inequality. To spot fake indigenous knowledge,  check if the source is reliable, whether the knowledge fits the cultural context, if the promotion is overly commercial, and whether ethical concerns are ignored. In the future, it will be a key research focus to use technical means to distinguish between genuine and fake indigenous knowledge.

\subsection{Harmonizing  the Technological Advancement with Computational Power}
In the realm of technological advancement and computational power balance, it is crucial to explore how to ensure the rational use and sustainability of computing resources while developing new technologies for safeguarding the sustainability of the indigenous knowledge supply chain. This includes investigating how technological innovations can enhance computational efficiency and reduce energy consumption \cite{10549890}. In addition, it involves weighing the trade-offs between large and small models in terms of computational power, energy consumption, and efficiency. It is also essential to consider how to apply these technologies across various fields to promote sustainable development and supply chain resilience. Meanwhile, attention must be paid to technological ethics to ensure that advancements do not bring about social risks or ethical dilemmas.

\subsection{Risk Governance on the Advanced AI Technology}
Future research should focus on addressing the failure risks of AI technologies applied in the indigenous knowledge supply chain, ensuring its security, efficiency, and sustainability.  Specifically, this involves establishing mechanisms for identifying and assessing ethical risks, enhancing ethical training for developers, and embedding ethical design from the outset to mitigate AI ``hallucination" issues \cite{Bi2024,Zhangai2025}. In addition, it is crucial to develop detection and defense tools for AI and supply chain failures, as well as corresponding supply chain financial risks due to the potential knowledge value exchanges, assess the direct and societal impacts of AI, and refine governance mechanisms that match the risk levels.

\subsection{Resilience Governance on Indigenous Knowledge Supply Chain}
The resilience governance of the indigenous knowledge supply chain is a crucial direction for future development. Currently, supply chains face dual risks of attack and failure: On the one hand, supply chain attacks are becoming increasingly complex, with attackers implanting malicious code by tampering with software, hardware, or service components \cite{10352334,Wang2023Adversarial}, Advanced Persistent Threat (APT) \cite{10517754}, and even employing social engineering for more covert attacks. On the other hand, supply chain failures may stem from technical faults or management loopholes, e.g., the lack of explainability in AI systems, data errors, model failures, or inadequate risk management mechanisms.
To address these challenges, future efforts should focus on two main areas: technology and management. Technologically, AI-driven tools for fault prediction, anomaly detection, and trend analysis can be utilized to identify potential risks in advance \cite{BI2023107172}. Managerially, comprehensive risk assessment, emergency response, and liability allocation mechanisms should be established to enhance the transparency and traceability of the supply chain. By doing so, it is expected that the resilience of the indigenous knowledge supply chain can be effectively enhanced, ensuring its security, efficiency, and sustainability in complex environments.

\subsection{Preventing Blockchain Failures in Indigenous Knowledge Supply Chain}
Preventing blockchain failures is a crucial direction for the future development of the indigenous knowledge supply chain \cite{blockchain_fail}. The unique nature of the indigenous knowledge supply chain demands even higher standards for addressing these issues in blockchain technology. First, performance issues such as slow transaction speeds and high resource consumption can hinder the efficient circulation and protection of indigenous knowledge, necessitating the optimization of consensus mechanisms and the adoption of lightweight blockchain technologies. Second, storage problems like insufficient data storage and backup may lead to the loss or irrecoverability of indigenous knowledge, which can be addressed through tiered storage strategies and multi-node backup solutions. Third, security issues such as smart contract vulnerabilities and poor key management can threaten the integrity and intellectual property rights of indigenous knowledge, requiring rigorous auditing and robust key management systems. Finally, privacy concerns like excessive data transparency and inadequate access control may leak sensitive information of indigenous communities, which can be mitigated through privacy protection technologies and hierarchical access control mechanisms. These measures not only ensure the stability and reliability of blockchain technology but also safeguard the indigenous knowledge supply chain in complex environments, ensuring its security, efficiency, and sustainability while respecting and protecting indigenous cultures and intellectual property rights.

\subsection{Converging Indigenous Knowledge with Classical Algorithms}
When adopting indigenous-inspired technical solutions, it is unwise to completely discard classical approaches. Instead, the efficient convergence of indigenous-inspired solutions with traditional methods will be a crucial direction for future research. Indigenous knowledge systems emphasize cultural sensitivity, transparency, and interpretability. Integrating these principles into classical algorithms not only enhances their applicability in specific cultural contexts but also boosts their transparency and credibility. By optimizing computational resources and dynamically adjusting parameters, the operational efficiency of algorithms can be further improved. In addition, incorporating concepts of data sovereignty and privacy protection strengthens the security of algorithms. This convergence is not only conducive to technological optimization but also promotes the protection and transmission of cultural heritage, offering new pathways for sustainable development.

\section{Conclusion}
This paper aims to address the crucial issue of protecting and transmitting indigenous knowledge on a global scale. By revealing the multifaceted challenges that currently impede the effective preservation and dissemination of indigenous knowledge, we have laid the groundwork for a more sustainable and inclusive approach. The proposed sustainable and trustworthy knowledge supply chain framework for indigenous knowledge serves as a comprehensive blueprint, integrating various stakeholders and processes to ensure the longevity and vitality of this invaluable cultural heritage.

Our review of existing technological solutions has highlighted both the progress made and the significant gaps that remain. While current technologies offer promising avenues for the protection and dissemination of indigenous knowledge, they often fall short in addressing the unique complexities and diverse contexts in which this knowledge exists. To bridge these gaps, we have introduced cutting-edge technologies that hold the potential to revolutionize the way indigenous knowledge is managed and shared. These technologies, including blockchain, AI, and digital twins, offer enhanced security, adaptability, and efficiency, thereby addressing many of the existing challenges.

Furthermore, this article has explored how the proposed framework can be practically applied to overcome real-world obstacles in the protection and transmission of indigenous knowledge. Leveraging advanced technologies, we can create more resilient and responsive supply chain ecosystems that not only safeguard indigenous knowledge but also empower indigenous communities to take greater ownership of their cultural heritage. The future research applications discussed in this article underscore the potential for continued innovation and adaptation, ensuring that these technologies remain relevant and effective in the face of evolving challenges.

At last, by addressing open issues and providing a detailed analysis, this article aims to guide future research and practice in the protection and transmission of indigenous knowledge. The promising research directions outlined here emphasize the need for interdisciplinary collaboration, community engagement, and technological innovation. We hope that this work will inspire further exploration and development, ultimately contributing to a more equitable and sustainable future for indigenous knowledge worldwide.
\bibliography{sample}

\begin{thebibliography}{100}
\urlstyle{rm}
\expandafter\ifx\csname url\endcsname\relax
  \def\url#1{\texttt{#1}}\fi
\expandafter\ifx\csname urlprefix\endcsname\relax\def\urlprefix{URL }\fi
\expandafter\ifx\csname doiprefix\endcsname\relax\def\doiprefix{DOI: }\fi
\providecommand{\bibinfo}[2]{#2}
\providecommand{\eprint}[2][]{\url{#2}}

\bibitem{un_2024}
\bibinfo{author}{{United Nations}}.
\newblock \bibinfo{title}{{UN}: Indigenous peoples have the right to 'voluntary
  isolation'}.
\newblock
  \bibinfo{howpublished}{\url{https://news.un.org/zh/story/2024/08/1130551}}
  (\bibinfo{year}{2024}).
\newblock \bibinfo{note}{Accessed: 2025-03-05}.

\bibitem{Hobson2010}
\bibinfo{author}{Hobson, J.}, \bibinfo{author}{Lowe, K.},
  \bibinfo{author}{Poetsch, S.} \& \bibinfo{author}{Walsh, M.}
\newblock \emph{\bibinfo{title}{Re-awakening Languages: Theory and Practice in
  the Revitalisation of Australia's Indigenous Languages}}
  (\bibinfo{publisher}{Sydney University Press}, \bibinfo{address}{Sydney},
  \bibinfo{year}{2010}).

\bibitem{Bromham2021GlobalPO}
\bibinfo{author}{Bromham, L.} \emph{et~al.}
\newblock \bibinfo{journal}{\bibinfo{title}{Global predictors of language
  endangerment and the future of linguistic diversity}}.
\newblock {\emph{\JournalTitle{Nature Ecology \& Evolution}}}
  \textbf{\bibinfo{volume}{6}}, \bibinfo{pages}{163 -- 173}
  (\bibinfo{year}{2022}).

\bibitem{NationalIndigenousLanguagesReport2020}
\bibinfo{author}{{Australian Government Department of Infrastructure,
  Transport, Regional Development and Communications, Australian Institute of
  Aboriginal and Torres Strait Islander Studies (AIATSIS), Australian National
  University}}.
\newblock \bibinfo{title}{{National Indigenous Languages Report}}.
\newblock \bibinfo{type}{Tech. Rep.} \bibinfo{number}{{Published 16th Aug
  2020}}, \bibinfo{institution}{{Office for the Arts}},
  \bibinfo{address}{{Australia}} (\bibinfo{year}{2020}).

\bibitem{aiatsis_nils3}
\bibinfo{author}{{AIATSIS}}.
\newblock \bibinfo{title}{{Introduction to the Third National Indigenous
  Languages Survey (NILS3)}}.
\newblock
  \bibinfo{howpublished}{\url{https://aiatsis.gov.au/third-national-indigenous-languages-survey-online/introduction}}.
\newblock \bibinfo{note}{Accessed: 2025-03-25}.

\bibitem{Stuart2017PigmentCI}
\bibinfo{author}{Stuart, B.~H.} \& \bibinfo{author}{Thomas, P.~S.}
\newblock \bibinfo{journal}{\bibinfo{title}{Pigment characterisation in
  {Australian} rock art: a review of modern instrumental methods of analysis}}.
\newblock {\emph{\JournalTitle{Heritage Science}}}
  \textbf{\bibinfo{volume}{5}}, \bibinfo{pages}{1--6} (\bibinfo{year}{2017}).

\bibitem{nmaFirstRockArt}
\bibinfo{author}{{National Museum of Australia}}.
\newblock \bibinfo{title}{{First rock art}}.
\newblock
  \bibinfo{howpublished}{{\url{https://www.nma.gov.au/defining-moments/resources/first-rock-art}}}.
\newblock \bibinfo{note}{{Accessed: 2025-03-20}}.

\bibitem{firstaustraliansepisode1}
\bibinfo{author}{{National Film and Sound Archive of Australia}}.
\newblock \bibinfo{title}{{First Australians: They Have Come to Stay}}.
\newblock
  \bibinfo{howpublished}{{\url{https://aso.gov.au/titles/documentaries/first-australians-episode-1/}}}.
\newblock \bibinfo{note}{{Accessed: 2024-10-21}}.

\bibitem{nmaSonglines}
\bibinfo{author}{{National Museum of Australia}}.
\newblock \bibinfo{title}{{Songlines: the foundational Australian story}}.
\newblock
  \bibinfo{howpublished}{{\url{https://www.nma.gov.au/audio/songlines-tracking-the-seven-sisters/transcripts/songlines-the-foundational-australian-story}}}.
\newblock \bibinfo{note}{{Accessed: 2025-03-25}}.

\bibitem{BaiYanYing201206009}
\bibinfo{author}{Bai, Y.} \emph{et~al.}
\newblock \bibinfo{journal}{\bibinfo{title}{Agricultural production in hani
  rice terraces system and related threats — a case study of zuofu and mitian
  villages in honghe county, china}}.
\newblock {\emph{\JournalTitle{Chinese Journal of Eco-Agriculture}}}
  \textbf{\bibinfo{volume}{20}}, \bibinfo{pages}{698--702}
  (\bibinfo{year}{2012}).

\bibitem{Dai}
\bibinfo{author}{Zhang, H.} \& \bibinfo{author}{Lei, T.}
\newblock \bibinfo{journal}{\bibinfo{title}{An ecological anthropological study
  of the traditional rice farming system of the dai people}}.
\newblock {\emph{\JournalTitle{Social Sciences in Yunnan}}}
  \bibinfo{pages}{155--161} (\bibinfo{year}{2018}).

\bibitem{Deang}
\bibinfo{author}{Li, Q.}
\newblock \bibinfo{journal}{\bibinfo{title}{Disaster early warning and the
  phenological calendar in de'ang agricultural activities}}.
\newblock {\emph{\JournalTitle{Journal of Southwest Minzu University(Humanities
  and Social Sciences Edition)}}} \textbf{\bibinfo{volume}{34}},
  \bibinfo{pages}{16--20} (\bibinfo{year}{2013}).

\bibitem{XNZS201510002}
\bibinfo{author}{Li, Y.}
\newblock \bibinfo{journal}{\bibinfo{title}{Ethnic traditional knowledge and
  disaster prevention and mitigation: A discussion on the disaster prevention
  and mitigation functions in yunnan minority cultures}}.
\newblock {\emph{\JournalTitle{Journal of Southwest Minzu University(Humanities
  and Social Sciences Edition)}}} \textbf{\bibinfo{volume}{36}},
  \bibinfo{pages}{1--6} (\bibinfo{year}{2015}).

\bibitem{cv3unkdf4397r3mvoukg}
\bibinfo{author}{{CNN}}.
\newblock \bibinfo{title}{Los angeles wildfires: Critical threat of new fires
  ends, but crews are still working to contain major blazes}.
\newblock
  \bibinfo{howpublished}{\url{https://edition.cnn.com/weather/live-news/fire-los-angeles-california-palisades-ventura-eaton-01-15-25-hnk/index.html}}.
\newblock \bibinfo{note}{Accessed: 2025-03-05}.

\bibitem{Natalie}
\bibinfo{author}{Ban, N.~C.} \emph{et~al.}
\newblock \bibinfo{journal}{\bibinfo{title}{Incorporate indigenous perspectives
  for impactful research and effective management}}.
\newblock {\emph{\JournalTitle{Nature Ecology \& Evolution}}}
  \textbf{\bibinfo{volume}{2}}, \bibinfo{pages}{1680–1683}
  (\bibinfo{year}{2018}).

\bibitem{WIPO1}
\bibinfo{author}{{WIPO}}.
\newblock \bibinfo{title}{Climate action and sustainability: Indigenous peoples
  are part of the solution}.
\newblock
  \bibinfo{howpublished}{\url{https://www.wipo.int/web/wipo-magazine/articles/climate-action-and-sustainability-indigenous-peoples-are-part-of-the-solution-41345}}.
\newblock \bibinfo{note}{Accessed: 2025-04-20}.

\bibitem{un_2024_1}
\bibinfo{author}{{United Nations}}.
\newblock \bibinfo{title}{Building resilient communities through indigenous
  consultation}.
\newblock
  \bibinfo{howpublished}{\url{https://www.un.org/en/academic-impact/we-are-indigenous-building-resilient-communities-through-indigenous-consultation}}.
\newblock \bibinfo{note}{Accessed: 2025-04-20}.

\bibitem{Antonelli2023IndigenousKI}
\bibinfo{author}{Antonelli, A.}
\newblock \bibinfo{journal}{\bibinfo{title}{Indigenous knowledge is key to
  sustainable food systems}}.
\newblock {\emph{\JournalTitle{Nature}}} \textbf{\bibinfo{volume}{613}},
  \bibinfo{pages}{239--242} (\bibinfo{year}{2023}).

\bibitem{Revive2023}
\bibinfo{author}{{Australian Government National Cultural Policy}}.
\newblock \bibinfo{title}{{Revive: A Place with All the Stories, A Story for
  Every Place}}.
\newblock
  \bibinfo{howpublished}{\url{https://www.arts.gov.au/what-we-do/new-national-cultural-policy}}.
\newblock \bibinfo{note}{Accessed: 2025-03-20}.

\bibitem{un2019indigenous}
\bibinfo{author}{{United Nations}}.
\newblock \bibinfo{title}{Indigenous people’s traditional knowledge must be
  preserved, valued globally, speakers stress as permanent forum opens annual
  session}.
\newblock
  \bibinfo{howpublished}{\url{https://press.un.org/en/2019/hr5431.doc.htm}}.
\newblock \bibinfo{note}{Accessed: 2024-09-24}.

\bibitem{colonization_impact}
\bibinfo{author}{Cusick, J.}
\newblock \bibinfo{title}{Impact of colonization on indigenous peoples’
  culture}.
\newblock
  \bibinfo{howpublished}{\url{https://opentextbc.ca/peersupport/chapter/impact-of-colonization-on-indigenous-peoples-culture}}.
\newblock \bibinfo{note}{Accessed: 2025-3-21}.

\bibitem{cvf2e2rduqb6sudn8ju0}
\bibinfo{author}{Mitchells, M.}
\newblock \bibinfo{title}{A silent crisis: Understanding why indigenous
  languages are disappearing}.
\newblock \bibinfo{note}{Accessed: 2025-03-05}.

\bibitem{Chen2023EvaluatingSS}
\bibinfo{author}{Chen, C.-C.}, \bibinfo{author}{Chen, W.},
  \bibinfo{author}{Zevallos, R.} \& \bibinfo{author}{Ortega, J.~E.}
\newblock \bibinfo{title}{Evaluating self-supervised speech representations for
  indigenous {American} languages}.
\newblock In \emph{\bibinfo{booktitle}{International Conference on Language
  Resources and Evaluation}} (\bibinfo{year}{2023}).

\bibitem{TonePahHote2022TheCO}
\bibinfo{author}{Tone-Pah-Hote, T.} \& \bibinfo{author}{Redvers, N.}
\newblock \bibinfo{journal}{\bibinfo{title}{The commercialization of
  biospecimens from indigenous peoples: A scoping review of benefit-sharing}}.
\newblock {\emph{\JournalTitle{Frontiers in Medicine}}}
  \textbf{\bibinfo{volume}{9}} (\bibinfo{year}{2022}).

\bibitem{Oyelude}
\bibinfo{author}{Oyelude, A.}
\newblock \bibinfo{journal}{\bibinfo{title}{Indigenous knowledge preservation
  as a sign of respect for culture: concerns of libraries, archives and
  museums}}.
\newblock {\emph{\JournalTitle{Insights the UKSG journal}}}
  \textbf{\bibinfo{volume}{36}} (\bibinfo{year}{2023}).

\bibitem{Orozco2013TheRO}
\bibinfo{author}{Orozco, D.} \& \bibinfo{author}{Poonamallee, L.}
\newblock \bibinfo{journal}{\bibinfo{title}{The role of ethics in the
  commercialization of indigenous knowledge}}.
\newblock {\emph{\JournalTitle{Journal of Business Ethics}}}
  \textbf{\bibinfo{volume}{119}}, \bibinfo{pages}{275 -- 286}
  (\bibinfo{year}{2013}).

\bibitem{traditional2024}
\bibinfo{author}{{United Nations}}.
\newblock \bibinfo{title}{Traditional knowledge – an answer to the most
  pressing global problems?}
\newblock
  \bibinfo{howpublished}{\url{https://www.un.org/en/desa/traditional-knowledge---answer-most-pressing-global-problems}}.
\newblock \bibinfo{note}{Accessed: 2025-03-19}.

\bibitem{Vijayan2022}
\bibinfo{author}{Vijayan, D.} \emph{et~al.}
\newblock \bibinfo{journal}{\bibinfo{title}{Indigenous knowledge in food system
  transformations}}.
\newblock {\emph{\JournalTitle{Communications Earth \& Environment}}}
  \textbf{\bibinfo{volume}{3}}, \bibinfo{pages}{1--7} (\bibinfo{year}{2022}).

\bibitem{lekhi2024loss}
\bibinfo{author}{Lekhi, B.}
\newblock \bibinfo{title}{Loss of traditional knowledge is due to lack of
  documentation}.
\newblock
  \bibinfo{howpublished}{\url{https://www.culturalsurvival.org/publications/cultural-survival-quarterly/loss-traditional-knowledge-due-lack-documentation}}.
\newblock \bibinfo{note}{Accessed: 2025-03-19}.

\bibitem{care_principles2023}
\bibinfo{author}{Jennings, L.} \emph{et~al.}
\newblock \bibinfo{journal}{\bibinfo{title}{Applying the ‘care principles for
  indigenous data governance’ to ecology and biodiversity research}}.
\newblock {\emph{\JournalTitle{Nature Ecology \& Evolution}}}
  \textbf{\bibinfo{volume}{7}}, \bibinfo{pages}{1547 -- 1551}
  (\bibinfo{year}{2023}).

\bibitem{science2022indigenous}
\bibinfo{author}{Sidik, S.~M.} \emph{et~al.}
\newblock \bibinfo{journal}{\bibinfo{title}{Weaving indigenous knowledge into
  the scientific method}}.
\newblock {\emph{\JournalTitle{Nature}}} \textbf{\bibinfo{volume}{601}},
  \bibinfo{pages}{285--287} (\bibinfo{year}{2022}).

\bibitem{Berkes2022}
\bibinfo{author}{Berkes, F.}
\newblock \bibinfo{journal}{\bibinfo{title}{Indigenous knowledge systems: A
  critical assessment}}.
\newblock {\emph{\JournalTitle{Ecology and Society}}}
  \textbf{\bibinfo{volume}{27}}, \bibinfo{pages}{30} (\bibinfo{year}{2022}).

\bibitem{MACKEY20222626}
\bibinfo{author}{Mackey, T.~K.} \emph{et~al.}
\newblock \bibinfo{journal}{\bibinfo{title}{Establishing a blockchain-enabled
  indigenous data sovereignty framework for genomic data}}.
\newblock {\emph{\JournalTitle{Cell}}} \textbf{\bibinfo{volume}{185}},
  \bibinfo{pages}{2626--2631} (\bibinfo{year}{2022}).

\bibitem{Consortium1}
\bibinfo{author}{Zheng, P.} \emph{et~al.}
\newblock \bibinfo{journal}{\bibinfo{title}{Meepo: {Multiple} execution
  environments per organization in sharded consortium blockchain}}.
\newblock {\emph{\JournalTitle{IEEE Journal on Selected Areas in
  Communications}}} \textbf{\bibinfo{volume}{40}}, \bibinfo{pages}{3562--3574}
  (\bibinfo{year}{2022}).

\bibitem{Cahyawijaya202414899}
\bibinfo{author}{Cahyawijaya, S.} \emph{et~al.}
\newblock \bibinfo{title}{Cendol: Open instruction-tuned generative large
  language models for {Indonesian} languages}.
\newblock vol.~\bibinfo{volume}{1}, \bibinfo{pages}{14899 – 14914}
  (\bibinfo{year}{2024}).

\bibitem{George2020463}
\bibinfo{author}{George, S.~K.}, \bibinfo{author}{Jagathy~Raj, V.} \&
  \bibinfo{author}{Gopalan, S.~K.}
\newblock \bibinfo{journal}{\bibinfo{title}{Personalized news media extraction
  and archival framework with news ordering and localization}}.
\newblock {\emph{\JournalTitle{Advances in Intelligent Systems and Computing}}}
  \textbf{\bibinfo{volume}{933}}, \bibinfo{pages}{463 – 471}
  (\bibinfo{year}{2020}).

\bibitem{Soosay2024379}
\bibinfo{author}{Soosay, T.} \& \bibinfo{author}{George, S.}
\newblock \bibinfo{title}{Two-eyed seeing: Vr learning with indigenous
  relevance}.
\newblock \bibinfo{pages}{379 – 384} (\bibinfo{year}{2024}).

\bibitem{Bai2024670}
\bibinfo{author}{Bai, Y.}, \bibinfo{author}{Li, X.}, \bibinfo{author}{Feng,
  Y.}, \bibinfo{author}{Liu, M.} \& \bibinfo{author}{Chen, C.}
\newblock \bibinfo{journal}{\bibinfo{title}{Preserving traditional systems:
  Identification of agricultural heritage areas based on agro-biodiversity}}.
\newblock {\emph{\JournalTitle{Plants People Planet}}}
  \textbf{\bibinfo{volume}{6}}, \bibinfo{pages}{670 – 682}
  (\bibinfo{year}{2024}).

\bibitem{Abed2025}
\bibinfo{author}{Abed, N.}, \bibinfo{author}{Murgun, R.},
  \bibinfo{author}{Deldari, A.}, \bibinfo{author}{Sankarannair, S.} \&
  \bibinfo{author}{Ramesh, M.~V.}
\newblock \bibinfo{journal}{\bibinfo{title}{Iot and ai-driven solutions for
  human-wildlife conflict: Advancing sustainable agriculture and biodiversity
  conservation}}.
\newblock {\emph{\JournalTitle{Smart Agricultural Technology}}}
  \textbf{\bibinfo{volume}{10}} (\bibinfo{year}{2025}).

\bibitem{Scoville202130}
\bibinfo{author}{Scoville, C.}, \bibinfo{author}{Chapman, M.},
  \bibinfo{author}{Amironesei, R.} \& \bibinfo{author}{Boettiger, C.}
\newblock \bibinfo{journal}{\bibinfo{title}{Algorithmic conservation in a
  changing climate}}.
\newblock {\emph{\JournalTitle{Current Opinion in Environmental
  Sustainability}}} \textbf{\bibinfo{volume}{51}}, \bibinfo{pages}{30 – 35}
  (\bibinfo{year}{2021}).

\bibitem{Ahire2023241}
\bibinfo{author}{Ahire, P.~R.}, \bibinfo{author}{Hanchate, R.} \&
  \bibinfo{author}{Varadarajan, V.}
\newblock \emph{\bibinfo{title}{Indigenous knowledge in smart agriculture}}
  (\bibinfo{year}{2023}).

\bibitem{SMPC1}
\bibinfo{author}{Morales, D.}, \bibinfo{author}{Agudo, I.} \&
  \bibinfo{author}{Lopez, J.}
\newblock \bibinfo{journal}{\bibinfo{title}{Toward a framework for
  cost-effective and publicly verifiable confidential computations in
  blockchain}}.
\newblock {\emph{\JournalTitle{IEEE Communications Magazine}}}
  \textbf{\bibinfo{volume}{63}}, \bibinfo{pages}{96 – 102}
  (\bibinfo{year}{2025}).

\bibitem{SMPC2}
\bibinfo{author}{El~Mestari, S.~Z.}, \bibinfo{author}{Lenzini, G.} \&
  \bibinfo{author}{Demirci, H.}
\newblock \bibinfo{journal}{\bibinfo{title}{Preserving data privacy in machine
  learning systems}}.
\newblock {\emph{\JournalTitle{Computers and Security}}}
  \textbf{\bibinfo{volume}{137}} (\bibinfo{year}{2024}).

\bibitem{huang2024survey}
\bibinfo{author}{Huang, D.}, \bibinfo{author}{Yan, C.}, \bibinfo{author}{Li,
  Q.} \& \bibinfo{author}{Peng, X.}
\newblock \bibinfo{journal}{\bibinfo{title}{A survey on multimodal large
  language models}}.
\newblock {\emph{\JournalTitle{arXiv preprint arXiv:2306.13549}}}
  (\bibinfo{year}{2024}).
\newblock \bibinfo{note}{Available: https://arxiv.org/pdf/2306.13549.pdf}.

\bibitem{MLLM1}
\bibinfo{author}{Han, S.~C.}, \bibinfo{author}{Cao, F.}, \bibinfo{author}{Poon,
  J.} \& \bibinfo{author}{Navigli, R.}
\newblock \bibinfo{title}{Multimodal large language models and tunings: Vision,
  language, sensors, audio, and beyond}.
\newblock \bibinfo{pages}{11294 – 11295} (\bibinfo{year}{2024}).

\bibitem{MLLM2}
\bibinfo{author}{Li, Z.} \emph{et~al.}
\newblock \bibinfo{title}{Enhancing advanced visual reasoning ability of large
  language models}.
\newblock \bibinfo{pages}{1915 – 1929} (\bibinfo{year}{2024}).

\bibitem{MLLM3}
\bibinfo{author}{Luo, Y.} \emph{et~al.}
\newblock \bibinfo{journal}{\bibinfo{title}{Biomedgpt: An open multimodal large
  language model for biomedicine}}.
\newblock {\emph{\JournalTitle{IEEE Journal of Biomedical and Health
  Informatics}}}  (\bibinfo{year}{2024}).

\bibitem{MLLM4}
\bibinfo{author}{Takahashi, R.}, \bibinfo{author}{Saito, N.},
  \bibinfo{author}{Maeda, K.}, \bibinfo{author}{Ogawa, T.} \&
  \bibinfo{author}{Haseyama, M.}
\newblock \bibinfo{title}{Personalized visual emotion classification via
  in-context learning in multimodal llm}.
\newblock \bibinfo{pages}{1167 – 1168} (\bibinfo{year}{2024}).

\bibitem{MLLM5}
\bibinfo{author}{Liu, F.} \emph{et~al.}
\newblock \bibinfo{journal}{\bibinfo{title}{A medical multimodal large language
  model for future pandemics}}.
\newblock {\emph{\JournalTitle{npj Digital Medicine}}}
  \textbf{\bibinfo{volume}{6}} (\bibinfo{year}{2023}).

\bibitem{bai2023qwen}
\bibinfo{author}{Bai, J.} \emph{et~al.}
\newblock \bibinfo{journal}{\bibinfo{title}{Qwen-vl: A versatile
  vision-language model for understanding, localization, text reading, and
  beyond}}.
\newblock {\emph{\JournalTitle{arXiv preprint arXiv:2308.12966}}}
  (\bibinfo{year}{2023}).
\newblock \bibinfo{note}{Available: https://arxiv.org/abs/2308.12966}.

\bibitem{xiao2024comprehensive}
\bibinfo{author}{Xiao, H.} \emph{et~al.}
\newblock \bibinfo{journal}{\bibinfo{title}{A comprehensive survey on medical
  large language models and multimodal large language models}}.
\newblock {\emph{\JournalTitle{arXiv preprint arXiv:2405.08603}}}
  (\bibinfo{year}{2024}).
\newblock \bibinfo{note}{Available:
  https://www.x-mol.com/paper/1790931362730094592/t}.

\bibitem{EC1}
\bibinfo{author}{Pioli, L.}, \bibinfo{author}{De~Macedo, D. D.~J.},
  \bibinfo{author}{Costa, D.~G.} \& \bibinfo{author}{Dantas, M. A.~R.}
\newblock \bibinfo{journal}{\bibinfo{title}{Intelligent edge-powered data
  reduction: A systematic literature review}}.
\newblock {\emph{\JournalTitle{ACM Computing Surveys}}}
  \textbf{\bibinfo{volume}{56}} (\bibinfo{year}{2024}).

\bibitem{EC2}
\bibinfo{author}{Baranwal, G.}, \bibinfo{author}{Kumar, D.} \&
  \bibinfo{author}{Vidyarthi, D.~P.}
\newblock \bibinfo{journal}{\bibinfo{title}{Blockchain based resource
  allocation in cloud and distributed edge computing: A survey}}.
\newblock {\emph{\JournalTitle{Computer Communications}}}
  \textbf{\bibinfo{volume}{209}}, \bibinfo{pages}{469 – 498}
  (\bibinfo{year}{2023}).

\bibitem{FL1}
\bibinfo{author}{Dembani, R.} \emph{et~al.}
\newblock \bibinfo{journal}{\bibinfo{title}{Agricultural data privacy and
  federated learning: A review of challenges and opportunities}}.
\newblock {\emph{\JournalTitle{Computers and Electronics in Agriculture}}}
  \textbf{\bibinfo{volume}{232}} (\bibinfo{year}{2025}).

\bibitem{FL2}
\bibinfo{author}{Mazid, A.}, \bibinfo{author}{Kirmani, S.},
  \bibinfo{author}{Abid, M.} \& \bibinfo{author}{Pawar, V.}
\newblock \bibinfo{journal}{\bibinfo{title}{A secure and efficient framework
  for internet of medical things through blockchain driven customized federated
  learning}}.
\newblock {\emph{\JournalTitle{Cluster Computing}}}
  \textbf{\bibinfo{volume}{28}} (\bibinfo{year}{2025}).

\bibitem{yuan2023Amplitude}
\bibinfo{author}{Yuan, X.} \emph{et~al.}
\newblock \bibinfo{journal}{\bibinfo{title}{Amplitude-varying perturbation for
  balancing privacy and utility in federated learning}}.
\newblock {\emph{\JournalTitle{IEEE Transactions on Information Forensics and
  Security}}} \textbf{\bibinfo{volume}{18}}, \bibinfo{pages}{1884--1897}
  (\bibinfo{year}{2023}).

\bibitem{Nan2024WWW}
\bibinfo{author}{Wu, N.}, \bibinfo{author}{Yuan, X.}, \bibinfo{author}{Wang,
  S.}, \bibinfo{author}{Hu, H.} \& \bibinfo{author}{Xue, M.}
\newblock \bibinfo{title}{Cardinality counting in ``alcatraz'': A privacy-aware
  federated learning approach}.
\newblock In \emph{\bibinfo{booktitle}{Proceedings of the ACM Web Conference
  2024}}, WWW '24, \bibinfo{pages}{3076–3084}
  (\bibinfo{publisher}{Association for Computing Machinery},
  \bibinfo{address}{New York, NY, USA}, \bibinfo{year}{2024}).

\bibitem{Li2025A}
\bibinfo{author}{Li, K.} \emph{et~al.}
\newblock \bibinfo{journal}{\bibinfo{title}{A novel framework of
  horizontal-vertical hybrid federated learning for edgeiot}}.
\newblock {\emph{\JournalTitle{IEEE Networking Letters}}} \bibinfo{pages}{1--1}
  (\bibinfo{year}{2025}).

\bibitem{Hu2024OFDMA}
\bibinfo{author}{Hu, S.} \emph{et~al.}
\newblock \bibinfo{journal}{\bibinfo{title}{{OFDMA}-{F²L}: Federated learning
  with flexible aggregation over an ofdma air interface}}.
\newblock {\emph{\JournalTitle{IEEE Transactions on Wireless Communications}}}
  \textbf{\bibinfo{volume}{23}}, \bibinfo{pages}{6793--6807}
  (\bibinfo{year}{2024}).

\bibitem{10948161}
\bibinfo{author}{Hu, S.} \emph{et~al.}
\newblock \bibinfo{journal}{\bibinfo{title}{Differentially private wireless
  federated learning with integrated sensing and communication}}.
\newblock {\emph{\JournalTitle{IEEE Transactions on Wireless Communications}}}
  \bibinfo{pages}{1--1} (\bibinfo{year}{2025}).

\bibitem{DP1}
\bibinfo{author}{Orlowski, A.} \& \bibinfo{author}{Loh, W.}
\newblock \bibinfo{journal}{\bibinfo{title}{Data autonomy and privacy in the
  smart home: the case for a privacy smart home meta-assistant}}.
\newblock {\emph{\JournalTitle{AI and Society}}}  (\bibinfo{year}{2025}).

\bibitem{DP2}
\bibinfo{author}{Xi, X.}, \bibinfo{author}{Zhang, Y.} \& \bibinfo{author}{Goh,
  M.}
\newblock \bibinfo{journal}{\bibinfo{title}{Consumer data collection strategies
  in two-sided platforms: The role of data ownership assignment and privacy
  concerns}}.
\newblock {\emph{\JournalTitle{International Journal of Production Economics}}}
  \textbf{\bibinfo{volume}{280}} (\bibinfo{year}{2025}).

\bibitem{Shan2023Preserving}
\bibinfo{author}{Shan, B.} \emph{et~al.}
\newblock \bibinfo{journal}{\bibinfo{title}{Preserving the privacy of latent
  information for graph-structured data}}.
\newblock {\emph{\JournalTitle{IEEE Transactions on Information Forensics and
  Security}}} \textbf{\bibinfo{volume}{18}}, \bibinfo{pages}{5041--5055}
  (\bibinfo{year}{2023}).

\bibitem{You2024Novel}
\bibinfo{author}{You, F.}, \bibinfo{author}{Yuan, X.}, \bibinfo{author}{Ni, W.}
  \& \bibinfo{author}{Jamalipour, A.}
\newblock \bibinfo{journal}{\bibinfo{title}{A novel privacy-preserving
  incentive mechanism for multi-access edge computing}}.
\newblock {\emph{\JournalTitle{IEEE Transactions on Cognitive Communications
  and Networking}}} \textbf{\bibinfo{volume}{10}}, \bibinfo{pages}{1928--1943}
  (\bibinfo{year}{2024}).

\bibitem{El-Beltagy202426}
\bibinfo{author}{El-Beltagy, A.~S.}
\newblock \emph{\bibinfo{title}{Navigation through uncertainties: Part B:
  Agro-ecosystems affected by dynamic impact of climate change}}
  (\bibinfo{year}{2024}).

\bibitem{Nyetanyane20203}
\bibinfo{author}{Nyetanyane, J.} \& \bibinfo{author}{Masinde, M.}
\newblock \bibinfo{journal}{\bibinfo{title}{Integration of indigenous
  knowledge, climate data, satellite imagery and machine learning to optimize
  cropping decisions by small-scale farmers. a case study of umgungundlovu
  district municipality, south africa}}.
\newblock {\emph{\JournalTitle{Lecture Notes of the Institute for Computer
  Sciences, Social-Informatics and Telecommunications Engineering, LNICST}}}
  \textbf{\bibinfo{volume}{321 LNICST}}, \bibinfo{pages}{3 – 19}
  (\bibinfo{year}{2020}).

\bibitem{Caudill2024}
\bibinfo{author}{Caudill, C.~M.}, \bibinfo{author}{Pulsifer, P.~L.},
  \bibinfo{author}{Thumbadoo, R.~V.} \& \bibinfo{author}{Taylor, D. R.~F.}
\newblock \bibinfo{journal}{\bibinfo{title}{Meeting the challenges of the un
  sustainable development goals through holistic systems thinking and applied
  geospatial ethics}}.
\newblock {\emph{\JournalTitle{ISPRS International Journal of
  Geo-Information}}} \textbf{\bibinfo{volume}{13}} (\bibinfo{year}{2024}).

\bibitem{Buitendag2024}
\bibinfo{author}{Buitendag, A.} \& \bibinfo{author}{Hattingh, F.}
\newblock \bibinfo{journal}{\bibinfo{title}{A smart agricultural knowledge
  management framework to support emergent farmers in developmental settings}}.
\newblock {\emph{\JournalTitle{Interdisciplinary Journal of Information,
  Knowledge, and Management}}} \textbf{\bibinfo{volume}{19}}
  (\bibinfo{year}{2024}).

\bibitem{Thothela2023}
\bibinfo{author}{Thothela, N.~P.}, \bibinfo{author}{Markus, E.},
  \bibinfo{author}{Masinde, M.} \& \bibinfo{author}{Abu-Mahfouz, A.}
\newblock \bibinfo{title}{A framework for an intelligent agro-climate decision
  support system for small-scale farmers in swayimane} (\bibinfo{year}{2023}).

\bibitem{Ahmed2024}
\bibinfo{author}{Ahmed, Z.}, \bibinfo{author}{Gui, D.},
  \bibinfo{author}{Abd-Elmabod, S.~K.}, \bibinfo{author}{Murtaza, G.} \&
  \bibinfo{author}{Ali, S.}
\newblock \bibinfo{journal}{\bibinfo{title}{An overview of global
  desertification control efforts: Key challenges and overarching solutions}}.
\newblock {\emph{\JournalTitle{Soil Use and Management}}}
  \textbf{\bibinfo{volume}{40}} (\bibinfo{year}{2024}).

\bibitem{Robinson2022}
\bibinfo{author}{Robinson, C.~J.} \emph{et~al.}
\newblock \bibinfo{journal}{\bibinfo{title}{Coproduction mechanisms to weave
  indigenous knowledge, artificial intelligence, and technical data to enable
  indigenous-led adaptive decision making: lessons from australia’s joint
  managed kakadu national park}}.
\newblock {\emph{\JournalTitle{Ecology and Society}}}
  \textbf{\bibinfo{volume}{27}} (\bibinfo{year}{2022}).

\bibitem{Ramba2023}
\bibinfo{author}{Ramba, P.}, \bibinfo{author}{Mbele, M.} \&
  \bibinfo{author}{Masinde, M.}
\newblock \bibinfo{title}{Integration of forth industrial revolution
  technologies and indigenous knowledge in developing a smart and integrated
  pollution monitoring system} (\bibinfo{year}{2023}).

\bibitem{Charbel2020708}
\bibinfo{author}{Charbel, H.} \& \bibinfo{author}{Lobato, D.~L.}
\newblock \bibinfo{title}{Between signal and noise : a trans-climatic approach
  in decoding and recoding autonomous ecologies}.
\newblock vol.~\bibinfo{volume}{1}, \bibinfo{pages}{708 – 718}
  (\bibinfo{year}{2020}).

\bibitem{Richard-Bollans2023}
\bibinfo{author}{Richard-Bollans, A.} \emph{et~al.}
\newblock \bibinfo{journal}{\bibinfo{title}{Machine learning enhances
  prediction of plants as potential sources of antimalarials}}.
\newblock {\emph{\JournalTitle{Frontiers in Plant Science}}}
  \textbf{\bibinfo{volume}{14}} (\bibinfo{year}{2023}).

\bibitem{Munn20241673}
\bibinfo{author}{Munn, L.}
\newblock \bibinfo{journal}{\bibinfo{title}{The five tests: designing and
  evaluating ai according to indigenous māori principles}}.
\newblock {\emph{\JournalTitle{AI and Society}}} \textbf{\bibinfo{volume}{39}},
  \bibinfo{pages}{1673 – 1681} (\bibinfo{year}{2024}).

\bibitem{GNN1}
\bibinfo{author}{Sun, H.}, \bibinfo{author}{Tu, Z.}, \bibinfo{author}{Sui, D.},
  \bibinfo{author}{Zhang, B.} \& \bibinfo{author}{Xu, X.}
\newblock \bibinfo{journal}{\bibinfo{title}{A federated social recommendation
  approach with enhanced hypergraph neural network}}.
\newblock {\emph{\JournalTitle{ACM Transactions on Intelligent Systems and
  Technology}}} \textbf{\bibinfo{volume}{16}} (\bibinfo{year}{2024}).

\bibitem{GNN2}
\bibinfo{author}{Qian, L.}, \bibinfo{author}{Cao, W.} \& \bibinfo{author}{Chen,
  L.}
\newblock \bibinfo{journal}{\bibinfo{title}{Influence of artificial
  intelligence on higher education reform and talent cultivation in the digital
  intelligence era}}.
\newblock {\emph{\JournalTitle{Scientific reports}}}
  \textbf{\bibinfo{volume}{15}}, \bibinfo{pages}{6047} (\bibinfo{year}{2025}).

\bibitem{GNN3}
\bibinfo{author}{Mishra, R.} \& \bibinfo{author}{Shridevi, S.}
\newblock \bibinfo{journal}{\bibinfo{title}{Knowledge graph driven medicine
  recommendation system using graph neural networks on longitudinal medical
  records}}.
\newblock {\emph{\JournalTitle{Scientific Reports}}}
  \textbf{\bibinfo{volume}{14}} (\bibinfo{year}{2024}).

\bibitem{10518175}
\bibinfo{author}{Li, K.} \emph{et~al.}
\newblock \bibinfo{journal}{\bibinfo{title}{Leverage variational graph
  representation for model poisoning on federated learning}}.
\newblock {\emph{\JournalTitle{IEEE Transactions on Neural Networks and
  Learning Systems}}} \textbf{\bibinfo{volume}{36}}, \bibinfo{pages}{116--128}
  (\bibinfo{year}{2025}).

\bibitem{KD1}
\bibinfo{author}{Wu, S.~X.} \emph{et~al.}
\newblock \bibinfo{journal}{\bibinfo{title}{Defender of privacy and fairness:
  Tiny but reversible generative model via mutually collaborative knowledge
  distillation}}.
\newblock {\emph{\JournalTitle{Neurocomputing}}} \textbf{\bibinfo{volume}{618}}
  (\bibinfo{year}{2025}).

\bibitem{KD2}
\bibinfo{author}{Xie, B.}, \bibinfo{author}{Xu, H.}, \bibinfo{author}{Seo, D.},
  \bibinfo{author}{Shin, D.} \& \bibinfo{author}{Cai, Z.}
\newblock \bibinfo{journal}{\bibinfo{title}{Kdgan: Knowledge distillation-based
  model copyright protection for secure and communication-efficient model
  publishing}}.
\newblock {\emph{\JournalTitle{IET Communications}}}
  \textbf{\bibinfo{volume}{18}}, \bibinfo{pages}{860 – 868}
  (\bibinfo{year}{2024}).

\bibitem{DT1}
\bibinfo{author}{Chaddad, A.} \& \bibinfo{author}{Jiang, Y.}
\newblock \bibinfo{journal}{\bibinfo{title}{Integrating technologies in the
  metaverse for enhanced healthcare and medical education}}.
\newblock {\emph{\JournalTitle{IEEE Transactions on Learning Technologies}}}
  (\bibinfo{year}{2025}).

\bibitem{DT2}
\bibinfo{author}{Hashash, O.} \emph{et~al.}
\newblock \bibinfo{journal}{\bibinfo{title}{The seven worlds and experiences of
  the wireless metaverse: Challenges and opportunities}}.
\newblock {\emph{\JournalTitle{IEEE Communications Magazine}}}
  \textbf{\bibinfo{volume}{63}}, \bibinfo{pages}{120 – 127}
  (\bibinfo{year}{2025}).

\bibitem{KG2}
\bibinfo{author}{Sellami, D.}, \bibinfo{author}{Inoubli, W.},
  \bibinfo{author}{Farah, I.~R.} \& \bibinfo{author}{Aridhi, S.}
\newblock \bibinfo{journal}{\bibinfo{title}{Knowledge graph representation
  learning: A comprehensive and experimental overview}}.
\newblock {\emph{\JournalTitle{Computer Science Review}}}
  \textbf{\bibinfo{volume}{56}} (\bibinfo{year}{2025}).

\bibitem{su14105881}
\bibinfo{author}{Bi, S.}, \bibinfo{author}{Ni, W.}, \bibinfo{author}{Jiang, Y.}
  \& \bibinfo{author}{Wang, X.}
\newblock \bibinfo{journal}{\bibinfo{title}{Novel recommendation-based approach
  for multidisciplinary development of future universities}}.
\newblock {\emph{\JournalTitle{Sustainability}}} \textbf{\bibinfo{volume}{14}}
  (\bibinfo{year}{2022}).

\bibitem{KG1}
\bibinfo{author}{Ratna, S.}, \bibinfo{author}{Singh, S.} \&
  \bibinfo{author}{Sharma, A.}
\newblock \bibinfo{journal}{\bibinfo{title}{An inclusive analysis for
  performance and efficiency of graph neural network models for node
  classification}}.
\newblock {\emph{\JournalTitle{Computer Science Review}}}
  \textbf{\bibinfo{volume}{56}} (\bibinfo{year}{2025}).

\bibitem{MAS1}
\bibinfo{author}{Luzolo, P.~H.}, \bibinfo{author}{Elrawashdeh, Z.},
  \bibinfo{author}{Tchappi, I.}, \bibinfo{author}{Galland, S.} \&
  \bibinfo{author}{Outay, F.}
\newblock \bibinfo{journal}{\bibinfo{title}{Combining multi-agent systems and
  artificial intelligence of things: Technical challenges and gains}}.
\newblock {\emph{\JournalTitle{Internet of Things (Netherlands)}}}
  \textbf{\bibinfo{volume}{28}} (\bibinfo{year}{2024}).

\bibitem{MAS2}
\bibinfo{author}{Nascimento, N.}, \bibinfo{author}{Alencar, P.} \&
  \bibinfo{author}{Cowan, D.}
\newblock \bibinfo{title}{Self-adaptive large language model (llm)-based
  multiagent systems}.
\newblock \bibinfo{pages}{104 – 109} (\bibinfo{year}{2023}).

\bibitem{You2024Privacy}
\bibinfo{author}{You, F.}, \bibinfo{author}{Yuan, X.}, \bibinfo{author}{Ni, W.}
  \& \bibinfo{author}{Jamalipour, A.}
\newblock \bibinfo{journal}{\bibinfo{title}{Privacy-preserving multi-agent deep
  reinforcement learning for effective resource auction in multi-access edge
  computing}}.
\newblock {\emph{\JournalTitle{IEEE Transactions on Cognitive Communications
  and Networking}}} \bibinfo{pages}{1--1} (\bibinfo{year}{2024}).

\bibitem{Meta1}
\bibinfo{author}{Sun, Q.}, \bibinfo{author}{Liu, Y.}, \bibinfo{author}{Chua,
  T.-S.} \& \bibinfo{author}{Schiele, B.}
\newblock \bibinfo{title}{Meta-transfer learning for few-shot learning}.
\newblock vol. \bibinfo{volume}{2019-June}, \bibinfo{pages}{403 – 412}
  (\bibinfo{year}{2019}).

\bibitem{Meta2}
\bibinfo{author}{Hospedales, T.}, \bibinfo{author}{Antoniou, A.},
  \bibinfo{author}{Micaelli, P.} \& \bibinfo{author}{Storkey, A.}
\newblock \bibinfo{journal}{\bibinfo{title}{Meta-learning in neural networks: A
  survey}}.
\newblock {\emph{\JournalTitle{IEEE Transactions on Pattern Analysis and
  Machine Intelligence}}} \textbf{\bibinfo{volume}{44}}, \bibinfo{pages}{5149
  – 5169} (\bibinfo{year}{2022}).

\bibitem{CA3}
\bibinfo{author}{Fiorentino, M.} \& \bibinfo{author}{Gabbard, J.~L.}
\newblock \bibinfo{journal}{\bibinfo{title}{Special issue on next-generation
  mixed-reality user experiences: Training, teaching, and learning}}.
\newblock {\emph{\JournalTitle{IEEE Computer Graphics and Applications}}}
  \textbf{\bibinfo{volume}{44}}, \bibinfo{pages}{11 – 12}
  (\bibinfo{year}{2024}).

\bibitem{CA2}
\bibinfo{author}{Fang, W.}, \bibinfo{author}{Chen, L.}, \bibinfo{author}{Han,
  L.} \& \bibinfo{author}{Ding, J.}
\newblock \bibinfo{journal}{\bibinfo{title}{Context-aware cognitive augmented
  reality assembly: Past, present, and future}}.
\newblock {\emph{\JournalTitle{Journal of Industrial Information Integration}}}
  \textbf{\bibinfo{volume}{44}} (\bibinfo{year}{2025}).

\bibitem{CA1}
\bibinfo{author}{Hsu, K.-C.} \& \bibinfo{author}{Liu, G.-Z.}
\newblock \bibinfo{journal}{\bibinfo{title}{The construction of a theory-based
  augmented reality-featured context-aware ubiquitous learning facilitation
  framework for oral communication development}}.
\newblock {\emph{\JournalTitle{Journal of Computer Assisted Learning}}}
  \textbf{\bibinfo{volume}{39}}, \bibinfo{pages}{883 – 898}
  (\bibinfo{year}{2023}).

\bibitem{Ahmad2024123}
\bibinfo{author}{Ahmad, W.}, \bibinfo{author}{Sen, A.},
  \bibinfo{author}{Eesley, C.} \& \bibinfo{author}{Brynjolfsson, E.}
\newblock \bibinfo{journal}{\bibinfo{title}{Companies inadvertently fund online
  misinformation despite consumer backlash}}.
\newblock {\emph{\JournalTitle{Nature}}} \textbf{\bibinfo{volume}{630}},
  \bibinfo{pages}{123 – 131} (\bibinfo{year}{2024}).

\bibitem{10549890}
\bibinfo{author}{Argerich, M.~F.} \& \bibinfo{author}{Patiño-Martínez, M.}
\newblock \bibinfo{journal}{\bibinfo{title}{Measuring and improving the energy
  efficiency of large language models inference}}.
\newblock {\emph{\JournalTitle{IEEE Access}}} \textbf{\bibinfo{volume}{12}},
  \bibinfo{pages}{80194--80207} (\bibinfo{year}{2024}).

\bibitem{Bi2024}
\bibinfo{author}{Bi, S.} \emph{et~al.}
\newblock \bibinfo{journal}{\bibinfo{title}{Failure analysis in next-generation
  critical cellular communication infrastructures}}.
\newblock {\emph{\JournalTitle{arXiv}}}  (\bibinfo{year}{2024}).

\bibitem{Zhangai2025}
\bibinfo{author}{Zhang, R.} \emph{et~al.}
\newblock \bibinfo{journal}{\bibinfo{title}{Toward democratized generative {AI}
  in next-generation mobile edge networks}}.
\newblock {\emph{\JournalTitle{IEEE Network}}}  (\bibinfo{year}{2025}).

\bibitem{10352334}
\bibinfo{author}{Bi, S.} \emph{et~al.}
\newblock \bibinfo{journal}{\bibinfo{title}{Detection and mitigation of
  position spoofing attacks on cooperative uav swarm formations}}.
\newblock {\emph{\JournalTitle{IEEE Transactions on Information Forensics and
  Security}}} \textbf{\bibinfo{volume}{19}}, \bibinfo{pages}{1883--1895}
  (\bibinfo{year}{2024}).

\bibitem{Wang2023Adversarial}
\bibinfo{author}{Wang, Y.} \emph{et~al.}
\newblock \bibinfo{journal}{\bibinfo{title}{Adversarial attacks and defenses in
  machine learning-empowered communication systems and networks: A contemporary
  survey}}.
\newblock {\emph{\JournalTitle{IEEE Communications Surveys \& Tutorials}}}
  \textbf{\bibinfo{volume}{25}}, \bibinfo{pages}{2245--2298}
  (\bibinfo{year}{2023}).

\bibitem{10517754}
\bibinfo{author}{Aly, A.}, \bibinfo{author}{Iqbal, S.},
  \bibinfo{author}{Youssef, A.} \& \bibinfo{author}{Mansour, E.}
\newblock \bibinfo{journal}{\bibinfo{title}{{MEGR-APT}: A memory-efficient
  {APT} hunting system based on attack representation learning}}.
\newblock {\emph{\JournalTitle{IEEE Transactions on Information Forensics and
  Security}}} \textbf{\bibinfo{volume}{19}}, \bibinfo{pages}{5257--5271}
  (\bibinfo{year}{2024}).

\bibitem{BI2023107172}
\bibinfo{author}{Bi, S.} \emph{et~al.}
\newblock \bibinfo{journal}{\bibinfo{title}{A comprehensive survey on
  applications of {AI} technologies to failure analysis of industrial
  systems}}.
\newblock {\emph{\JournalTitle{Engineering Failure Analysis}}}
  \textbf{\bibinfo{volume}{148}}, \bibinfo{pages}{107172}
  (\bibinfo{year}{2023}).

\bibitem{blockchain_fail}
\bibinfo{author}{Najati, I.}
\newblock \bibinfo{journal}{\bibinfo{title}{Exploring the failure factors of
  blockchain adopting projects: A case study of tradelens through the lens of
  commons theory}}.
\newblock {\emph{\JournalTitle{Frontiers in Blockchain}}}
  \textbf{\bibinfo{volume}{8}} (\bibinfo{year}{2025}).

\end{thebibliography}

\newpage

\end{document}